# PHYSICAL CONDITIONS IN ORION'S VEIL

N. P. Abel,[1] C. L. Brogan,[2] G. J. Ferland,[1] C. R. O'Dell,[3] G. Shaw,[1] & T. H. Troland[1]

[1]University of Kentucky, Department of Physics and Astronomy, Lexington, KY 40506; npabel2@uky.edu, gary@pa.uky.edu, gshaw@pa.uky.edu, troland@pa.uky.edu

[2]Institute for Astronomy, 640 North A'ohoku Place, Hilo, HI 96720 and JCMT Fellow; cbrogan@ifa.hawaii.edu

[3]Department of Physics and Astronomy, Vanderbilt University, Box 1807-B, Nashville, TN 37235; cr.odell@vanderbilt.edu

## Abstract

Orion's veil consists of several layers of largely neutral gas lying between us and the main ionizing stars of the Orion nebula. It is visible in 21cm H I absorption and in optical and UV absorption lines of H I and other species. Toward $\theta^1$ Ori C, the veil has two remarkable properties, high magnetic field (≈ 100 μG) and a surprising lack of $H_2$ given its total column density. Here we compute photoionization models of the veil to establish its gas density and its distance from $\theta^1$ Ori C. We use a greatly improved model of the $H_2$ molecule that determines level populations in $10^5$ rotational/vibrational levels and provides improved estimates of $H_2$ destruction via the Lyman-Werner bands. Our best fit photoionization models place the veil 1-3 pc in front of the star at a density of $10^3$-$10^4$ cm$^{-3}$. Magnetic energy dominates the energy of non-thermal motions in at least one of the 21cm H I velocity components. Therefore, the veil is the first interstellar environment where magnetic dominance appears to exist. We find that the low ratio of $H_2/H^0$ ($< 10^{-4}$) is a consequence of high UV flux incident upon the veil due to its proximity to the Trapezium stars and the absence of small grains in the region.

# 1 Introduction

The Orion Complex is one of the best-studied regions of active star formation. These studies include a wide variety of observations of the magnetic field on many angular scales (e.g. van der Werf et al. 1993, Heiles 1997). Star formation is controlled by a complex and dynamic interplay between gravitational, thermal, and magnetic forces. Magnetic effects, in particular, can be complex. Yet they must be considered if we are to achieve a comprehensive understanding of the star formation process.

The Orion region consists of ionized ($H^+$), atomic ($H^0$), and molecular gas. The H II region is mainly the illuminated face of the background molecular cloud, Orion Molecular Cloud 1 (OMC1; see the comprehensive review by O'Dell 2001). It is heated and ionized by the light of the Trapezium cluster, with the single O6 V star $\theta^1$ Ori C (Walborn & Panek, 1984) providing most of the energy. The "veil" is the layer of largely atomic gas that lies between the Trapezium and us. Reddening within the H II region itself is small (O'Dell 2002). Also, optical obscuration toward the nebula correlates well with $H^0$ column density in the veil (O'Dell & Yusef-Zadeh 2000). Therefore, nearly all obscuration toward the H II region occurs within the veil. Moreover, studies of stellar extinction as a function of distance establish that the veil is associated with the Orion complex (Goudis, 1982).

A wealth of observational data reveals physical properties of the veil. In particular, the veil is one of the few interstellar regions where we have accurate *maps* of the line of sight component of the local magnetic field (Troland et al. 1989). These magnetic field maps are derived from Zeeman splitting of the H I 21 cm line, seen in absorption against the H II region continuum emission. Other physical properties of the veil such as density, temperature, and level of ionization, can be determined by combining optical and UV absorption line measurements with photoionization calculations. Altogether, these data present a unique opportunity to test modern theories of the physical state of the interstellar medium in star-forming regions.

Section 2 summarizes what is known about the physical state of gas in Orion's veil. Section 3 confronts these observations with simulations of the expected thermal and ionization equilibrium. In Section 4 we discuss the significance of the model results to magnetic energy equipartition in the veil, and in Section 5 we summarize our conclusions.

# 2 Observed properties of the veil

## 2.1 Column densities and extinction

Observations of absorption lines reveal the column densities of a number of atomic and molecular species in the veil. Shuping & Snow (1997; henceforth referred to as SS97) derive the atomic hydrogen column density, $N(H^0)$, from IUE observations of the Ly$\alpha$ line. Their value of $N(H^0) = (4 \pm 1) \times 10^{21}$ cm$^{-2}$ is a factor of ~4 larger than the previous OAO-2 and Copernicus measurements (Savage & Jenkins 1972; Savage et al. 1977). Column densities of other species were also measured by SS97. Among the most important for this study are excited states of $C^0$, $O^0$, and $Si^+$. Since the populations of more than one excited state have been measured for each, they can be used as density indicators. The atomic and singly ionized column densities of C, Mg, and S were also measured, and this information constrains the ionization state of the veil. We assume that the physical properties of the gas within the veil are driven by the radiation emitted by the nearby Trapezium stars.

Reddening and extinction to the Trapezium stars and over the entire H II region have been measured by a number of authors (Bohlin & Savage 1981; Cardelli, Clayton, & Mathis 1989; O'Dell & Yusef-Zadeh 2000). The color excess E(B-V) = 0.32 mag, and the extinction is $A_V$ = 1.6 mag. Therefore the ratio of total to selective extinction $R$ = 5.0, significantly larger than the average value of 3.1 in the diffuse ISM. The high value of $R$ suggests that the reddening is caused by grains with a larger than normal size distribution. The measured visual extinction and the SS97 value for $N(H^0)$ yield a ratio, $A_V/N(H^0)$, of $(4\pm1)\,x10^{-22}$ mag cm$^2$. This value is slightly lower than the average ratio in the diffuse interstellar medium ($5.3 \times 10^{-22}$ mag cm$^2$), and it is also consistent with larger than normal grains. If any ionized hydrogen is present in the veil, the ratio of visual extinction to total hydrogen column density will be even smaller.

The line of sight to the Trapezium has a surprisingly low fraction of molecular gas. Copernicus observations established that $N(H_2) < 3.5x10^{17}$ cm$^{-2}$ (Savage et al. 1977). This value implies $N(H_2)/N(H^0) \lesssim 10^{-4}$ (other regions within the Orion complex are molecular, of course). However, observed correlations between $N(H_2)/N(H^0)$ and $A_V$ (Savage et al. 1977) suggest that the veil should be predominantly molecular. This anomaly has long been thought to arise from the veil's abnormal grain properties (SS97), although a high UV flux has also been suggested (Savage et al. 1977). Larger grains are less effective at producing $H_2$ since the grain surface area per unit grain mass is smaller. Also, large grains are less effective in shielding $H_2$ from photodissociation by the Lyman-Werner bands because of the smaller projected area per unit mass. The observed $N(H^0)$ and the absence of $H_2$ in the veil (toward the Trapezium stars) will be important constraints on the model calculations presented in Section 3.

## 2.2 H I 21 cm velocity components

Van de Werf & Goss (1989) reported at least three separate velocity components in the 21cm H I absorption lines seen against the background H II region. More recent observations of these same 21 cm lines were carried out by Troland et al. (2004; in preparation) with the Very Large Array (VLA) of the NRAO[1] and are presented in Figure 1. This line profile reveals two relatively narrow components toward the Trapezium and possibly a broader underlying component. For the purposes of this analysis, we fitted the Troland et al. H I optical depth profile with just two Gaussian components. Inclusion of the underlying broad component does not alter the analysis below in any significant way. Of the two components in the fit, the "narrow" component has ($V_{LSR}$, $\Delta V_{FWHM}$) = 5.3 and 2.0 km $s^{-1}$, the "wide" component has ($V_{LSR}$, $\Delta V_{FWHM}$) = 1.3 and 3.8 km $s^{-1}$. (Add 18.1 km $s^{-1}$ to convert LSR to heliocentric velocities.) If the line widths result from thermal broadening alone, kinetic temperatures are 87 and 315 K in the narrow and wide components, respectively. However, 21 cm H I lines are almost always broader than thermal, indicating the presence of supersonic motions (Heiles & Troland, 2003). Therefore, actual kinetic temperatures may be significantly less.

[1] The National Radio Astronomy Observatory (NRAO) is a facility of the National Science Foundation operated under cooperative agreement by Associated Universities, Inc.

## 2.3 Optical velocity components

For reference, a schematic geometry of the central regions of Orion, based on O'Dell (1993), is shown in Figure 2.

Line profiles can be used to reveal kinematics of the various regions. The SS97 UV data had insufficient spectral resolution to resolve lines into individual components. However, the much higher resolution optical study performed by O'Dell et al. (1993) detected many components in the Ca II absorption line toward the Trapezium stars ($\theta^1$ Ori A, B, C, and D) and toward $\theta^2$ Ori. The Ca II profiles for Ori A-D have an absorption component in the same velocity range as the H I 21 cm absorption. (Ca II and 21 cm H I lines are shown in Figure 1.) Yet the Ca II profile has other components not seen in 21 cm H I absorption.

The lack of a strong correlation between Ca II and 21 cm H I absorption profiles may have a simple explanation. The $Ca^+$ column densities measured by O'Dell et al. (1993) show little variation, ranging from 10.7 to 12.9x$10^{11}$ $cm^{-2}$. For a gas phase Ca / H abundance of 2x$10^{-8}$ (Baldwin et al. 1991, hereafter BFM), the associated $H^0$ column density is $6 \times 10^{19}$ $cm^{-2}$. This value is less than 2% of $N(H^0)$ measured by SS97. Therefore, $Ca^+$ is a trace constituent within the veil. Photoionization calculations presented later confirm this conclusion. We believe $H^0$ samples most of the matter in the veil, so we will focus on this species in our calculations.

## 2.4 Physical location, size, and geometry

Statistical arguments suggest that the veil is ~0.6 pc away from $\theta^1$ Ori C. This estimate is based on the 3-D distribution of stars in the Orion Nebula Cluster given by Hillenbrand & Hartmann (1998) and upon observations of the fraction of proplyds that are protected by the veil from photoionization and seen in silhouette (Bally et al. 2000). The uncertainty of this value is hard to estimate.

Another constraint on the veil location comes from the fact that the veil emission in Hβ is not prominent compared with the Hβ surface brightness of the background H II region (5x$10^{-12}$ erg $s^{-1}$ $cm^{-2}$ $arcsec^{-2}$, BFM 1991). This constraint requires the veil to be much further away from the Trapezium stars than the main photoionized region (~0.25 parsecs, Figure 2). Since the veil's position is poorly constrained by these arguments, we will treat it as a free parameter.

The geometric properties of the veil and its density are partly constrained by observational data. The veil subtends at least 10 arc minutes in the plane of the sky since it covers both M42 and M43 (van der Werf & Goss 1989). This dimension amounts to 1.5 pc at 500 pc. The thickness $L$ can be deduced from the SS97 $H^0$ column density if the atomic hydrogen density $n(H^0)$ is also known since $L \sim 4x10^{21} / n(H^0)$ cm. If the veil extends no more than ~1.5 pc in the plane of the sky, then this relation sets a lower limit to the density, $n(H^0) > 10^2$ $cm^{-3}$, since the veil is probably not cigar-shaped pointed in our direction. There is no real physical upper limit to the density. If $n(H^0) \geq 10^3$ $cm^{-3}$, then the veil is a sheet viewed face on. Such sheet-like structures are commonly seen in the diffuse ISM (Heiles & Troland 2003).

## 2.5 Harmonic mean temperature of $H^0$

The 21 cm H I and Lyα data establish the harmonic mean temperature of the veil. (See Section 3.3.) The 21 cm absorption measures $N(H^0)/T_{spin}$ because of the correction for stimulated emission (Spitzer 1978; equation 3-37). From the 21 cm data of van der Werf and Goss (1989), $N(H^0)/T_{spin} = 4.3x10^{19}$ $cm^{-2}$ $K^{-1}$ toward the Trapezium stars (integrated across the full line). This value, and the measured $N(H^0)$ from SS97, imply a harmonic mean spin temperature of 70-120K (corresponding to the lower and upper limits to $N(H^0)$ derived by SS97). However, the narrow H I component must have T < 87 K, the temperature equivalent of its full width.

Harmonic mean $H^0$ temperatures, estimated from Lyα and 21 cm data are only accurate if both data sample the same absorbing gas. This assumption need not be correct since the Lyα and 21 cm data sample very different projected areas through the veil (sub-arcsec for the UV vs ~15 arcsec for the radio). However, maps of optical extinction suggest it is fairly smooth across the veil (O'Dell & Yusef-Zadeh 2000), so the Lyα and 21 cm data likely sample comparable regions.

## 2.6 Magnetic properties

Using the Zeeman effect in the 21 cm absorption lines, Troland et al. (1989, 2004) measure line-of-sight magnetic field strengths $B_{los}$ independently in the narrow and wide H I velocity components. Across much of the veil, they find comparable values for the two components, typically -50 μG, where the minus sign indicates a field pointed toward the observer. Along the line of sight to the Trapezium, $B_{los} = -45 \pm 5$ and $-54 \pm 4$ μG for the narrow and wide components, respectively. For a statistical ensemble of randomly-directed fields of the same strength, the average total field strength is $2B_{los}$. Therefore, toward the Trapezium stars, we take the total field strength to be -100 μG in the veil.

Field strengths this high are almost exclusively associated with regions of massive star formation where $n > 10^3$-$10^4$ $cm^{-3}$ (Crutcher et al. 1999). They are never seen in diffuse, absorbing H I gas of the cold neutral medium (CNM) studied by Heiles & Troland (2003). Nor are they commonly found in molecular clouds associated with low and intermediate mass star formation (Crutcher et al. 1993, Troland et al. 1996). High field strengths in the veil almost certainly reflect the history of this layer, a history associated with a large, self-gravitating region of massive star formation rather than with more typical diffuse ISM.

Equipartition between the (non-thermal) kinetic energy in MHD waves and the energy in the magnetic field is thought to arise commonly in the ISM (Zweibel & McKee 1995). Also, such equipartition is consistent with observations of regions where magnetic field strengths have been measured (Myers & Goodman 1988; Crutcher 1999). If equipartition exists between magnetic field energy and the energy of non-thermal motions in the gas, then we can express the gas density in terms of the field strength $B$ and the contribution to the line width from non-thermal motions $\Delta v_{NT}$. Equating the magnetic field energy with the energy of non-thermal motions,

$$B^2/8\pi = \frac{1}{2}\rho \Delta v_{NT}^2 \ , \tag{1}$$

we find upon appropriate units conversion

$$n_{eq} = \left[ \frac{2.5B}{\Delta v_{NT}} \right]^2 \ [\mathrm{cm}^{-3}]. \tag{2}$$

Here $n_{eq}$ is the equipartition *proton* density $[n(H^0) + 2n(H_2)]$, $B$ is in μG, and $\Delta v_{NT}$ is the FWHM line width attributable to non-thermal motions in the gas (km s$^{-1}$). If $n / n_{eq} > 1$, non-thermal kinetic energy dominates. If $n / n_{eq} < 1$, magnetic energy dominates, and the field lines should be relatively uniform.

# 3 Photoionization calculations

## 3.1 The density – distance grid

Our goal is to deduce the density of the veil and its distance from the Trapezium stars by comparing photoionization models of this layer with observations. We used the development version of the spectral synthesis code Cloudy, last described by Ferland et al. (1999). Bottorff et al. (1998) and Armour et al. (1999) provide further details about this code.

Our veil models incorporate several important improvements in Cloudy. We use the improved grain physics described by van Hoof et al. (2001) which explicitly solves for the grain temperature and charge as a function of grain size. We divide the grain size distribution into ten size bins, and include grains that are composed of graphite and silicates. Our size distribution is designed to reproduce the flat UV extinction curve observed by Cardelli et al. (1989) towards

Orion. We will see below (Section 3.4 and Table 1) that the best-fit models predict about 5% of the hydrogen along the line of sight to the Trapezium is ionized and so is not seen in $H^+$ absorption line studies. This result, combined with the neutral column density measurements of SS97, leads to $A_V/N(H_{tot}) \approx 4\times10^{-22}$ mag cm$^2$, a ratio which is ~75% of that seen in the general ISM. We scaled our grain abundance to match this ratio.

We have implemented a complete model of the $H_2$ molecule into Cloudy. Energies and radiative transition probabilities for the 301 ro-vibrational levels within the ground electronic state $1s\ ^1\Sigma_g$ (denoted as X) are taken from Stancil (2002, private communication) and Wolnievicz et al. (1998). We have included the ro-vibrational levels within the lowest 6 electronic excited states that are coupled to the ground electronic state by permitted electronic transitions. Energies and radiative transition probabilities for excited electronic states are from Abgrall, Roueff, & Drira (2000). These electronic excited states are important because they determine the Solomon process, which destroys $H_2$ through the absorption of Lyman-Werner band photons from the ground electronic state followed by decays into the X continuum. These photo-excitations are also an indirect source of population of excited ro-vibration levels within X which decay to produce infrared emission lines with the selection rule $\delta J = 0, \pm 2$. We have taken photoionization cross sections for transitions into the continuum of the Lyman and Werner bands from Allison & Dalgarno (1969). Effects of cosmic rays and the x-ray continuum are included as well. $H_2$ can be formed either on dust grains in a cold dusty environment or from $H^-$ in a hot dust-free environment. The state specific ($v$ and $J$ resolved) formation rates of $H_2$ on grain surfaces and via $H^-$ route have been taken from Cazaux & Tielens (2002); Takahashi, Junko, & Uehara (2001); and Launay, Dourneuf, & Zeippen (1991). Line overlap and self-shielding are also considered. In addition, our calculations are designed to reproduce the observationally determined $H_2$ formation rate on grains in the diffuse ISM, determined by Jura (1974) to be ~$3x10^{-17}$ cm$^{-3}$ s$^{-1}$.

Other important input parameters to the models include gas-phase abundances and the adopted stellar ionizing continuum. We use gas-phase abundances observed in the Orion Nebula (Rubin et al. 1991; Osterbrock et al 1992, Baldwin et al. 1991). A few of the abundances by number are He / H = 0.095, C / H = $3.0x10^{-4}$, N / H = 7.0 $x10^{-5}$, O / H=4.0 $x10^{-4}$, Ne / H=$6.0x10^{-5}$ and Ar / H =3.0 $x10^{-6}$. However, all of the lightest thirty elements are included in our models. The stellar ionizing continuum is the modified Kurucz LTE atmosphere described by Rubin et al. (1991), which was modified to reproduce high-ionization lines seen in the H II region. We set the total number of ionizing photons emitting by $\theta^1$ Ori C equal to $10^{49.34}$ s$^{-1}$, a typical value for an O6 star (Osterbrock, 1989). The radiation field falls off as the inverse square of the separation from the star, so the star-veil separation becomes a model parameter. Finally, the ionization and thermal effects of background galactic cosmic rays are included using the ionization rate quoted by McKee (1999).

We computed a large grid of models in which two parameters were varied. These parameters are the veil's total hydrogen density, denoted by $n_H = n(H^+) + n(H^0) + 2n(H_2)$, and the distance between the side of the veil facing the trapezium and the trapezium itself, which is designated by *D*. For simplicity we assume a single constant-density layer. This is a reasonable assumption since the predicted gas temperatures depend upon the shielding of the layer from the radiation field, not on the details of how this shielding occurs. All calculated models have the total atomic hydrogen column density measured by SS97, log$[N(H^0)]$ = 21.6. From our results we were able to identify ranges of densities and distances that reproduce the observed features of the veil. A more thorough description of our results is described next.

## 3.2 Column densities, ionization structure, and depletion

SS97 derived column densities of many species in the veil in addition to $H^0$. These results, along with the upper limit to the $H_2$ column density, allow us to constrain the level of ionization, density, and to some extent the distance of the veil from the Trapezium cluster. Our primary focus here is on column densities for species with more than one stage of ionization. These include atomic and singly ionized C, Mg, and S. Column density ratios of different ion states of the same element determine the level of ionization in a way that does not depend on the abundance. These data determine the ionization parameter, the ratio of the flux of ionizing photons to the density. SS97 derived column densities of many more elements (such as Fe, Zn, and P), but they were detected in only one stage of ionization. Therefore, these column densities do not constrain the model since we could easily reproduce them by simply scaling the abundances. The column densities of excited state $O^0$, $C^0$, and $Si^+$ have also been measured. The ratio of column densities of excited to ground state determines the density if the density is less than the critical density of the species. If the density is known, then the measured ionization parameter can be used to determine the distance from the continuum source. The measured column densities are summarized in Table 1.

Model grids for species with more than one observed ionization or excitation state are shown in Figures 3 {A, B, and C} and 4 {A, B, C, D, and E}, respectively. Figure 3 shows our predicted ratios of $Mg^0/Mg^+$, $S^0/S^+$, and $C^0/C^+$. There is a large range of parameters that match the observations. All regions that produce $N(Mg^0)/N(Mg^+) < 10^{-3}$ are also consistent with the observed ratios $N(S^0)/N(S^+)$ and $N(C^0)/N(C^+)$. The parameter space with large distances and densities predict these species to be mostly atomic, and is therefore excluded. Figure 4 shows the predicted ratios of excited to ground state column densities for the five species measured by SS97. As expected, these ratios are very sensitive to the hydrogen density. Our calculation reproduce the N(CI*)/ N(CI) and N(CI**)/ N(CI) ratios to within a factor of two for densities >$10^3$ $cm^{-3}$. Observational uncertainties were not stated for these column densities. The N(OI*)/ N(OI) and N(OI**)/ N(O I) column density ratios are a better constraint in the models since, like Mg, 1σ error bars are given. The 1σ ranges are given in Table 1. The N(OI*)/ N(OI) ratio suggests n ~ $10^{3\pm0.4}$ $cm^{-3}$, while the N(OI**)/ N(OI) suggests a density closer to $10^{3.7\pm0.5}$ $cm^{-3}$. The N(SiII*)/ N(SiII) ratio gives $n = 10^{4.5\pm0.5}$. A density of $n$ ~ $10^3$ $cm^{-3}$ matches the observed column densities within 3σ.

Both the $H_2$ formation and destruction physics are included in a self-consistent manner. We show in the Appendix that the predicted $H_2$ column density is very sensitive to the method used in calculating the Solomon process. We also show in the Appendix the sensitivity of predicted $H_2$ column densities to the grain size distribution. We find that the lack of $H_2$ in the veil is due to the efficiency of the Trapezium stars in destroying $H_2$, and to the sensitivity of molecular hydrogen formation to the grain size distribution. The continuum radiation responsible for the Solomon process is mainly the high-energy range of the Balmer continuum, between 11.2 and 13.6 eV. The right side of Figure 5 gives $G_o$, the average interstellar flux between 6 and 13.6 eV as defined in Tielens & Hollenbach (1985), for the range of distances we consider. Its large value ($G_o > 10^3$) leads to most of our calculated $H_2$ column densities being well under the Copernicus limit of $N(H_2) < 3.5x10^{17}$ $cm^{-2}$. Generally, only in regions of parameter space of higher density further away from the ionizing star ($D$~ $10^{19.0}$ cm) does $G_o$ become small enough for a large amount of $H_2$ to form.

Figure 5 can be used to derive a constraint to the distance $D$ if the density $n_H$ is known. Section 3.4 shows that the range of densities that best reproduce observations is ~$10^3 < n_H < 10^4$ $cm^{-3}$. Since the $H_2$ column density must be less than the upper limit derived by Copernicus, $N(H_2) < 3.5x10^{17}$ $cm^{-2}$, our models constrain $D$ to be less than 3 parsecs.

## 3.3 The spin temperature and surface brightness

The temperature measured by the 21 cm/Lyα ratio is an $N(H^0)$ -weighted harmonic mean temperature, given by the relationship

$$\frac{1}{<T_{spin}>} = \frac{\int \frac{n}{T_{spin}} dr}{N(H^0)} . \qquad (3)$$

Predicted values are shown in Figure 6. Like the excited state column densities, the spin temperature is sensitive to the density. For our calculations, the primary heating and cooling mechanisms are grain photoionization and collisional excitation of heavy element fine structure lines respectively. The two most efficient coolants were the [C II] 158μ and the [O I] 63μ lines. For lower densities both [C II] and [O I] contribute to the cooling. However, for densities larger than the [C II] 158μ critical density of ~3x10$^3$ cm$^{-3}$, cooling becomes less efficient, and the heating rate increases with a power of density greater than the cooling rate. This explains the increase in temperature seen in Figure 6 for densities greater than 10$^3$ cm$^{-3}$. For the 1 σ range of spin temperature set by the observed 21 cm / Lyα limit, $70 < T_{spin} < 110$, our predictions agree with observations for densities between 10$^{3.5}$-10$^{4.5}$ cm$^{-3}$. The 3 σ range for the spin temperature, 50-200 K, allows densities greater than ~10$^2$ cm$^{-3}$.

Although the veil is known to be deficient in small grains (Section 2.1), we considered the possible effects of PAHs in the model calculations. Very small grains, or large molecules like PAHs, are known to be very efficient at heating gas. As a check on their effects, we ran one set of models with PAHs present at an abundance of $n(PAH) / n(H) = 3 \times 10^{-7}$. As expected, the PAHs raised our predicted temperatures by factors between 1.5 and 2 over those shown in Figure 6. The effect of higher temperatures can only be accommodated at lower densities. That is, the models predict slightly lower densities with the inclusion of PAHs. Therefore, conclusions in Section 4 based upon relatively low veil densities are not endangered by the effects of PAHs in the models.

Figure 7 shows the predicted Hβ surface brightness emitted by the ionized face of the veil. The contours are very nearly parallel, depending mainly on the distance, since the surface brightness in a recombination line is determined by the flux of ionizing photons striking the gas. Furthermore, the ionizing flux (hence, the Hβ surface brightness) is an inverse square function of the star-cloud distance. If we assure that the surface brightness in the veil is no more than half of that observed in the main ionization front, then a lower limit for the distance from the veil to $\theta^1$ Ori C is $D$~ 10$^{18.0}$ cm (~0.33pc).

A firmer lower limit on $D$ can be derived through the use of [N II] 6583Å emission, which must be occurring at the surface of the veil that faces $\theta^1$ Ori C. The difficulty is distinguishing veil emission from radiation from the main body of the nebula. An upper limit on the [N II] 6583Å veil emission can be placed by assuming the veil accounts for the red shoulder of the main nebular emission line. In this case the surface brightness of the veil is 1.5x10$^{-13}$ ergs cm$^{-2}$ s$^{-1}$ arcsec$^{-2}$ (or about -12.8 on a logarithmic scale), after correction for extinction (O'Dell & Yusef-Zadeh 2000). However, this is a generous upper limit since it is likely that much of the red shoulder is caused by red-shifted nebular light that is scattered by dust in the dense PDR that lies beyond the main ionization front of the nebula (O'Dell 2001).

Figure 8 shows the predicted [N II] surface brightness as a function of density and distance. All combinations of density and distance in Figure 8 with [N II] greater than -12.8 (corresponding to contours below the -12.8 contour in Figure 8) are excluded by observation. For densities less than 10$^4$ cm$^{-3}$, this constraint places a lower limit on the distance of ~10$^{18.5}$ cm (~1 parsecs). Our best models (Section 3.4) calculate $n_H$ to be between 10$^3$ and 10$^4$ cm$^{-3}$. Therefore, this upper limit upon $D$ applies. The limit $D > 1$ pc is about two times larger than the value deduced on statistical grounds (Section 2.4). Since our calculations best match observations for densities between $10^3$ and $10^4$ cm$^{-3}$, we are only concerned with the part of Figure 8 that constrains the distance to around 1 parsecs. As discussed below, our models with $D > 10^{18.5}$ cm also yield the best agreement with the rest of the observational data.

In Figure 9 we present contours of $H^+$ column density as a function of $D$ and $n_H$. For high density and large $D$, the amount of $H^+$ starts to drop off. For high density, the amount of $H^+$ starts to drop off due to a lower ionization parameter, defined as the dimensionless ratio of hydrogen ionizing flux to density. This behavior also affects the [N II] surface brightness at high density and large $D$ (Figure 8). Since nitrogen has a higher ionization potential than hydrogen, nitrogen will only be ionized in regions where hydrogen is also ionized. This leads to less $N^+$ in these regimes and therefore a lower surface brightness for the 6583 Å line.

## 3.4 A "best model" for the veil and its consequences

In this section we will define a "best model" that most accurately reproduces the observational features in the veil. Based on our calculations for the excited state abundance ratios and spin temperatures discussed in Section 3.2 & 3.3, a density between $10^3$-$10^4$ cm$^{-3}$ would reproduce all the observational data to within 3 $\sigma$. The distance of the veil, ~$10^{18.5}$ to $10^{19.0}$ cm (1 to 3 parsecs), is set by the observed upper limits on both the surface brightness of [N II] 6583Å and the column density of $H_2$. These are the broad limits to any successful model of the veil.

We derived a "best" model using the optimization methods that are part of Cloudy, starting with density and distance near the center of the allowed range but allowed to vary. The observables that we optimized include column densities of CI, CII, MgI, MgII, SI, SII, SiII, OI, SiII*, OI*, OI**, CI*, CI**, $H_2$, and the temperature $T_{spin}$. We used as confidence intervals the 1 $\sigma$ error estimates given by SS97. When these are not given a percent error of 20% was assumed. Where upper limits were given (as in the case of $C^+$, $S^0$, and $H_2$) all that was required in the calculation was that the predicted column densities be less than these upper limits. We then calculated a $\chi^2$ based on the formula:

$$\chi^2 = \left( \frac{F_{obs} - F_{pred}}{\sigma \times min\left(F_{obs}, F_{pred}\right)} \right)^2 \quad (4)$$

where $F_{obs}$ is the observed mean value, $F_{pred}$ is the predicted value generated by Cloudy, and $\sigma$ is the error in the observed value.

Tables 1 & 2 along with Figure 10 {a, b, and c} summarize our calculation that had the lowest $\chi^2$ and therefore represents our "best model" of the physical conditions in the veil. The distance is $10^{18.8\pm0.1}$cm (~2 parsecs) and hydrogen density is $n_H = \sim 10^{3.1\pm0.2}$ cm$^{-3}$, where the ranges correspond to densities and distances that yield a $\chi^2$ that are within a factor of two of the lowest value. The general conclusions drawn from our "best model" are representative of all models in this general range of the parametric diagrams.

Our optimal calculations predict column densities in unobserved stages of ionization. The majority of calcium in the veil is in the form of $Ca^{++}$ (see Table 2 & Figure 10b). Doubly ionized calcium has a closed shell and hence is undetectable in UV or optical radiation. Optical studies must work with $Ca^{+}$ which is a trace stage of ionization. It is for this reason that we take the 21 cm H I absorption line as the principal tracer of atomic gas (Section 2.3). We also predict that about 5% of the hydrogen in the veil is ionized.

Formalized optimization aside, how certain can we be that veil densities are as low as $n_H \sim 10^3$ cm$^{-3}$ (best model) and not $n_H \geq 10^4$ cm$^{-3}$ as required for magnetic equipartition in the narrow HI component (Section 4.2)? The best determined line ratios (therefore, the best diagnostics for determining density) are O I*/O I, O I**/O I, Si II*/Si II, and $Mg^0/Mg+$ (Section 3.2). Line ratios involving C and S either have undetermined observational errors or else errors too high to be useful (see Table 1). For our optimized density ($n_H = \sim 10^{3.1\pm0.2}$ cm$^{-3}$), the predicted O I*/O I ratio is within 1 $\sigma$ of observation. If $n_H \geq 10^4$ cm$^{-3}$, then the predicted ratio is only within 3 $\sigma$ of observation. Likewise, the predicted ratio $Mg^0/Mg^+$ at the optimized density lies within 1 $\sigma$ of observation, while the predicted density for $n_H \geq 10^4$ cm$^{-3}$ falls only within 3 $\sigma$ of observation. However, our models do not predict the Si II*/Si II ratio very well at either $n_H = \sim 10^3$ or $10^4$ cm$^{-3}$. For both densities, the predicted ratio barely falls within 3 $\sigma$ of observation. Also, the predicted ratio O I**/O I lies within 2 $\sigma$ of observation at either value for $n_H$. In general, our calculations show that densities $>10^4$ cm$^{-3}$ combined with distances greater than 1 parsec (the minimum distance allowed by the [N II] surface brightness measurement), will predict the veil to be both less ionized and more molecular than suggested by observation. However, the best case for $n_H = \sim 10^3$ cm$^{-3}$ as opposed to $10^4$ cm$^{-3}$ comes from the ratios O I*/O I and $Mg^0/Mg^+$ mentioned above.

Future observations of the line of sight towards $\theta^1$ Ori C done at higher spectral resolution could further increase our knowledge of the physical conditions in the veil. Observations in the UV, done at higher resolution than possible by IUE, would allow for the determination of the physical properties in each of the veil's components seen in H I 21 cm absorption. In a manner similar to what we have done in this work, excited state column densities along with column densities of elements seen in more than one ionization state could be calculated for each component in the veil. This information could then be used to derive the density, temperature, and level of ionization for each component in the veil. Additionally, observations of $H_2$ performed with greater sensitivity than the Copernicus mission could yield a more precise value for the column density of $H_2$ in the veil.

# 4 Implications for the magnetic field in the veil

## 4.1 Magnetic vs. non-thermal energies in the veil

Knowledge of the magnetic field strength, gas density, and non-thermal line width in the veil allows us to estimate the energetic importance of the magnetic field. The ratio of energy in non-thermal motions to energy in the field equals $n_H / n_{eq}$, the ratio of the actual gas density to the equilibrium density (Section 2.6, densities are proton densities, i.e. densities for $H^0$). Values of $n$ are inferred from our models where, for the veil, $n_H \approx n(H^0)$. In these models, higher $n$ is associated with higher $T_s$ over a wide range of veil distance $d$ (Figure 6). A lower limit to $n_{eq}$ comes from equation 2 assuming $\Delta v_{NT} = \Delta v_{obs}$, that is, the observed line width $\Delta v_{obs}$, has no thermal broadening. For non-zero temperature, $\Delta v_{NT} < \Delta v_{obs}$, and $n_{eq}$ is higher. Therefore, the question of magnetic energy dominance is addressed by estimating the ratio $n_H / n_{eq}$ as a function of $T_s$ over the allowable $T_s$ range for the veil. To do so, we make use of the model-determined $T_s$ - $n_H$ relationship of Figure 6.

## 4.2 The narrow HI component

We now consider the ratio $n_H / n_{eq}$ for the narrow H I component. The upper limit to $T_s$ is 87 K, set by $\Delta v_{obs}$. Although the harmonic mean temperature of the veil is 70-120 K (Section 2.5), the 21 cm and Ly$\alpha$ data set no lower limit upon $T_s$ for the narrow component alone. We have estimated the ratio $n_H / n_{eq}$ over the range $0 < T_s < 87$ K. To do so, we used $\Delta v_{obs} = 2$ km s$^{-1}$, $B = 100$ µG (Section 2.6), implying $n_{eq} \geq 1.6 \times 10^4$ cm$^{-3}$. We also used the $n_H$ - $T_s$ relationship in Figure 6 over the range $D = 10^{18.8 \pm 0.1}$ cm of our best models (Section 3.4). At all values of $T_s$, the ratio $n_H / n_{eq} < 0.025$, that is, magnetic energy is greater than the energy of non-thermal motions by at least a factor of 40. Even at the minimum possible $B = B_{los} \approx 50$ µG (Section 2.6), the magnetic energy dominates by an order of magnitude. Note that for $T_s < 87$ K Figure 6 allows for higher densities ($n_H > 10^4$ cm$^{-2}$) at larger values of $D$ ($> 10^{19}$ cm). However, this range of parameter space is excluded by the observed upper limit upon $N(H_2)$ as shown in Figure 5. That is, a high density veil located far from $\theta^1$ C Ori would become largely molecular, contrary to observations. The conclusion seems quite certain that magnetic fields dominate non-thermal motions in the narrow H I component.

It is possible that the field lies nearly along the line-of-sight with transverse MHD wave motions nearly in the plane of the sky. In such a case, the measured (line-of-sight) line width of the narrow component yields an underestimate of the actual non-thermal motions, hence, an underestimate of $n_H / n_{eq}$. We regard such an effect as very unlikely. A very high degree of alignment and MHD wave coherence would be necessary to reduce significantly the gas motions along the line-of-sight compared to those in the plane of the sky.

Such a pronounced dominance of magnetic field energy is unknown anywhere else in the interstellar medium. It is unexpected theoretically and unprecedented observationally (Section 2.6). Dominance by the magnetic field implies that the field should be rather uniform owing to tension in the magnetic field lines (e.g. Chandrasekhar & Fermi 1953). In fact, the magnetic field map across the veil reveals a rather uniform distribution of $B_{los}$ except in the northeast where $B_{los}$ increases substantially in the direction of the dark bay of obscuring material (Troland et al., 2004; in preparation). Magnetic dominance and a uniform field might arise if there is little input of mechanical energy to the gas or if the mechanical energy has damped out. Perhaps this is the case in the narrow component. Van der Werf & Goss (1989) identified this component with quiescent gas undisturbed by the H II region. However, Watson, Wiebe and Crutcher (2001) analyzed the statistics of variation of magnetic fields in the Troland et al. data set. They concluded that field variations across the veil were consistent with magnetic equipartition. Such a conclusion seems difficult to reconcile with the present results for the narrow component.

Turbulent line broadening is not necessarily related to MHD waves. For example, O'Dell, Peimbert & Peimbert (2003) argue that the anomalous broadening of 10 to 20 km / s, observed in optical emission lines from the H II region, is not an MHD effect. Conceivably, $H^0$ gas in the veil may also exhibit non-magnetic turbulent broadening. However, the veil is known to have a significant magnetic field. Therefore, non-thermal motions in the veil will inevitably be coupled to the field unless magnetic flux freezing is extremely inefficient. Therefore, our assumption that line broadening in the veil is magnetic seems inescapable.

## 4.3 The wide HI component

A similar assertion of magnetic dominance cannot be made for the wide H I component ($\Delta v_{obs}$ = 3.8 km s$^{-1}$). For this component $n_{eq} \geq 4 \times 10^3$ cm$^{-3}$ (equation 2 with B = 100 μG and $\Delta v_{NT} = \Delta v_{obs}$). This limit upon the equipartition density is comparable to the best fit model density of $10^{3.1}$ cm$^{-3}$, so $n_H / n_{eq} \approx 1$. In short, parameters for the wide H I component are consistent with magnetic equilibrium in this layer of the veil so they are consistent with the analysis of Watson et al. (2001).

# 5 Conclusions

We have combined radio, optical and UV data with an extensive series of photoionization models to define better the physical conditions in the veil of Orion. The models have been computed in a grid where the density of the veil and its distance from the principal source of ionizing photons, $\theta^1$ Ori C are treated as free parameters. This study has led to an improved understanding of the location of the veil, its density, its ionization structure (including the atomic and molecular hydrogen content) and the role of the magnetic field within it. Our principal conclusions are summarized below. These conclusions apply to the line-of-sight through the veil to the Trapezium stars, principally $\theta^1$ Ori C.

- Model calculations (Section 3) constrain the density in the veil to $\sim 10^3 \leq n_H \leq \sim 10^4$ cm$^{-3}$. The models also reveal a monotonic connection between density and temperature, higher temperatures are associated with higher densities for $n_H \leq 10^4$ cm$^{-3}$ (Figure 6). The best fit model has $n_H \sim 10^{3.1}$ cm$^{-3}$ at $T \sim 70$ K.
- Model calculations place the veil 1-3 parsecs away from the Trapezium, the best fit value is 2 pc. Also, measurements of the surface brightness of [N II] 6583Å require the veil to be at least 1 parsec from the Trapezium. These distances are about 2-5 times greater than the 0.6 pc value suggested by previous statistical analyses.
- Model calculations suggest $H_2$ is under abundant due to a combination of the high UV flux incident upon the veil and the larger-than-normal grains. Also, the predicted $H_2$ abundance is very sensitive to the details describing the dissociation and formation processes (Appendix). This means that straightforward approximations previously used in calculating the dissociation rate may not accurately predict $H_2$ abundances in at least some circumstances. In this study, we have used a greatly improved model for the $H_2$ molecule that incorporates over $10^5$ rotational / vibrational levels.
- The models confirm that $Ca^+$ is only a trace stage of ionization in the veil. Therefore, studies of Ca II optical absorption profiles may not be representative of the most of the matter in the veil. However, 21cm H I absorption traces all of the matter in the veil except the small fraction in the form of H+. Model calculations predict that about 5% of the hydrogen in the veil is ionized.
- There are two principal velocity components in the 21 cm H I absorption which we designate as the narrow and the wide components. Magnetic field measurements made via the Zeeman effect (Troland et al.,1989; 2004 in preparation) reveal the line-of-sight field strength $B_{los}$ across the veil. In the direction of the Trapezium stars, $B_{los} \sim -50$ μG for each of these components, with the total field strength statistically expected to be two times larger. We show from model estimates of density in the veil that the narrow H I component is strongly dominated by the magnetic field. That is, the field energy is much larger than the energy associated with non-thermal motions in the gas. The narrow H I component in the veil is the only interstellar environment known with this property.

- For the derived density of $10^{3.1}$ $cm^{-3}$ and observed column density of $N(H^0) \approx 10^{21.6}$ $cm^{-2}$, the physical thickness of the layer is $\approx$1 pc. The veil covers > 1.5 parsecs in the plane of the sky, its aspect ratio is >1:1.5, so is not necessarily a sheet. Were the density high enough to be in energy equipartition, $n \sim 10^4$ $cm^{-3}$, the veil would be a thin sheet, with an aspect ratio greater than 1:15.

Acknowledgements: We also would like to thank referee Dr. Mark Wolfire for his invaluable suggestions with drafts of this work. GJF thanks the NSF for support through AST 03-07720 and NASA for NAG5-12020. THT would like to acknowledge the support of the NSF grants 99-88341 and 03-07642. CRO's support was in part from the Space Telescope Science Institute's grant GO-9141.

# 7 Appendix – Sensitivity of N($H_2$) to Formation and Destruction Rates

In this section we show the extreme sensitivity that predicted molecular hydrogen column densities have to the formation and destruction rates.

In the interstellar medium, molecular hydrogen forms primarily through catalysis of two hydrogen atoms on grain surfaces. Destruction of $H_2$ occurs through photoabsorption in the Lyman-Werner bands. Calculating formation and destruction rates for the $H_2$ molecule can be computationally expensive, since the number of rotational/vibrational levels for $H_2$ is ~$10^5$. We will refer to the treatment of all levels of $H_2$ in calculating formation and destruction rates as using the "complete $H_2$ molecule", as in Section 3.2. The population of each level is determined by balancing processes that correspond to formation into it, destruction from it, and transitions into and out of it into other levels.

Often approximations to the formation and destruction rates are made, with the ultimate goal of increasing calculation speed while at the same time making predictions for $H_2$ abundances to the same level of accuracy as the complete $H_2$ molecule. Then the chemistry can be treated as a simple two-component system, and the balance written as

$$n\left(H_2\right)R_d = \frac{1}{2}n(H^0)nR_f \text{ cm}^{-3}\text{ s}^{-1} \quad (5)$$

where the photodestruction and grain formation rates are given by $R_d$ and $R_f$, the stoichiometric factor is ½, and $n(H^0)$ and $n$ are the atomic and neutral hydrogen densities [$n = 2n\left(H_2\right) + n\left(H^0\right)$].

## 7.1 Destruction Rates

Two widely used approximations to the destruction rates of $H_2$ come from Bertoldi & Draine (1996, henceforth referred to as BD96) and Tielens & Hollenbach (1985, referred to henceforth as TH85). While these approximations increase computational speed, the true dissociation rate may deviate significantly from simple analytical approximations if the UV field excites electrons to many different rotational/vibrational levels.

We tested how variations in these simple approximations to the destruction rates changed the predicted $H_2$ column density. First, in the case of destruction, we took the destruction rate given in the TH85 paper and scaled it in the following manner:

$$R_d = X \cdot 3.4 \cdot 10^{-11}\beta\left(\tau\right)G_0 e^{-2.5Av} \text{ s}^{-1} \quad (6)$$

Here $\beta(\tau)$ is a self-shielding factor that characterizes the amount of the continuum between 6 and 13.6eV that has been absorbed by the Lyman-Werner bands. As $H_2$ becomes more shielded from these photons, $\beta(\tau)$ will decrease. The parameter $G_0$ is the flux of UV photons between 6 and 13.6eV relative to the average interstellar value, with the exponential factor accounting for dust absorption of photons in the same range of energy. The scale factor $X$ is a free parameter that we use to test how the predicted $H_2$ column density varies with changes in the destruction rate. The BD96 destruction rate is very similar with the exception that self-shielding is unity for $H_2$ column densities less than $10^{14}$ $cm^{-2}$, which in terms of equation 5 means that $\beta(\tau)$ varies differently with depth .

As Figure 11 shows, scaling the simple dissociation rate by factors of < 10 can cause the predicted $H_2$ column density to change by several orders of magnitude. This sensitivity is due to the non-linear effects that changes in $H_2$ have on the self-shielding. Increasing the $H_2$ destruction rate will decrease the amount of $H_2$. Consequently, a decreased $H_2$ abundance will lead to less absorption of the Lyman-Werner bands (a higher $\beta(\tau)$), and hence less self shielding. This positive feedback mechanism produces the large changes in $H_2$ abundance for small changes to the scaling factor seen in Figure 11.

One would not expect either the TH85 or BD96 destruction rate to exactly match the $H_2$ destruction rate generated by Cloudy. Cloudy self-consistently determines the photodestruction rate for each line including attenuation by absorption, reemission by the gas, and line overlap. This is done for each depth into a cloud. Combining the use of a complete $H_2$ molecule with a self-consistent treatment of the radiation field gives the most physically realistic treatment of the Solomon process.

We compared the predicted $H_2$ abundance for three different treatments of the destruction rate, keeping all other parameters fixed. We took our best model and then changed the way we calculated the destruction rate to either that of TH85, BD96, or the complete $H_2$ molecule in Cloudy. Our results shown in Table 3 clearly demonstrate how approximations to the destruction rate can lead to dramatically different results. For the TH85 rate, the predicted column density was 4 orders of magnitude larger than either BD96 or the complete $H_2$ molecule. This difference suggests that treating the $H_2$ molecule as a two level atom causes the calculation to overcompensate for the effects of self-shielding, which lowers the calculated destruction rate. The lower destruction rate then causes an overabundance of $H_2$ through the same non-linear feedback that is seen in Figure 11. Because the BD96 approximation sets $\beta(\tau)$ equal to one for $N(H_2) < 10^{14}$ $cm^{-2}$, this approximation will lead to a slightly larger destruction rate and hence less $H_2$ formation. However, it appears that in general the BD96 dissociation rate approximates the Cloudy dissociation rate fairly well.

## 7.2 Formation Rates

In addition to testing the sensitivity of the predicted $H_2$ abundance to the destruction rate, we also tested the sensitivity of the $H_2$ abundance to the formation rate of $H_2$ by grain catalysis. TH85 use a formation rate given by:

$$R_f = X \cdot 6\times10^{-17}\left(T/300\right)^{0.5} S\left(T\right) \text{ cm}^3 \text{ s}^{-1} \qquad (7)$$

T is the temperature, and *S(T)* is the sticking coefficient taken from Hollenbach and McKee (1979). Again, we insert a scaling factor *X* to test how the $H_2$ abundance changes with formation rate.

Figure 11 also shows how scaling the formation rate affects the predicted $H_2$ abundance. Just like with the destruction rate, the final amount of $H_2$ is very sensitive to changes in the formation rate. Again, this is due to the non-linear relationship between the rate and the $H_2$ density. Because the $H_2$ fraction depends on the grain formation rate and *n* in equation 5, scaling the formation rate will make the $H_2$ abundance increase / decrease which in turn feeds back into equation 5 through *n*. This feedback quickly leads to large changes in $H_2$ formation for relatively small changes in X'. Since the formation rate is dependent on the treatment of the grain physics and the observationally determined rate coefficient (Jura, 1974), care must be taken to assure that both of these factors are determined to the highest possible precision.

## 7.3 Grain Size Distribution

We ran one final test, to check how the $H_2$ column density changed when different grain distributions were considered. We ran a series of models with a density of $10^3$ cm$^{-3}$, approximately corresponding to our best model density. We then varied $G_0$ from 1 to $10^6$ in increments of 0.5 dex. This calculation was performed for both an Orion grain distribution (absence of small grains) and a standard ISM grain distribution, designed to reproduce the standard interstellar extinction curve (R=3.1). This calculation, unlike the rest of our calculations, neglected hydrogen ionizing radiation. We also stopped this calculation at a total hydrogen column density of $4x10^{21}$ cm$^{-2}$, as opposed to just the atomic column density $N(H^0)$ in all other calculations.

As shown in Figure 12, the final $H_2$ column density can be very sensitive to the grain size distribution. For low values of $G_0$ both size distributions are effective in absorbing the $H_2$ dissociating continuum. This low dissociation rate allows all the available hydrogen to combine into $H_2$. As the value of $G_0$ is increased, a point is reached where, for a given size distribution and total hydrogen column density, the $H_2$ dissociating continuum is not effectively absorbed. The value of $G_0$ where this occurs will be smaller for grain size distributions that are weighted towards large grains. Larger grains provide less extinction per unit mass, which in turn keeps the dissociation rate large over a greater depth into the cloud.

Figure 12 shows that the critical value of $G_0$ where the dissociation rate remains large (i.e. self-shielding no longer occurs) is about $10^{2.5}$ for Orion grains and about $10^4$ for ISM grains. Over the range of $10^{2.5} < G_0 < 10^{4.5}$, differences in the predicted $H_2$ column density approaching 6 orders of magnitude can occur. This range of $G_0$ includes the value of about $10^4$ that we estimate for the veil. Therefore, the anomalous grain size distribution in the veil is capable of having a very important effect upon the $H_2$ abundance. For $G_0 > 10^{4.5}$, the sensitivity of $H_2$ to grain size is lessened. For these high values of $G_0$, neither size distribution is effective in absorbing the $H_2$ dissociation continuum, which keeps the dissociation rate high. The $H_2$ column density will remain small and, as can be seen in equation 5, scale inversely with the dissociation rate. The grain size distribution still affects the predicted $H_2$ abundance, because the larger grain sizes in Orion are less effective than the ISM grains in absorbing the UV continuum. The Orion grains also have less surface area per unit mass of hydrogen, which lowers the $H_2$ formation rate on grains relative to the ISM grain size distribution. These combined effects cause the predicted $H_2$ column density to be ~1.5 dex lower for our Orion grain size distribution for $G_0 > 10^{4.5}$.

The importance of using accurate formation and destruction rates is not new, Browning et al. (2003) also demonstrate this principal in a different context. However, it is not widely know that small deviations from the true formation and destruction rates can cause $H_2$ abundance predictions to vary by large amounts, at least for conditions similar to the veil. One must be sure that any assumptions or approximations to the formation and destruction rates do not significantly affect the final predictions. The best way to do this is to check the approximation versus the more refined, computationally expensive calculation. In the case of this paper, the TH85 destruction rates often led to results that differed significantly from observation. Our “best model” would not be allowed by observation if we had used the TH85 destruction rate. It was not until we calculated a destruction rate determined by using the “complete $H_2$ molecule” that we obtained results that were consistent with observational data.

# 8 Tables

Table 1

Best Fit Parameters for the Veil

| Parameter | Model[2] | Observed 1 $\sigma$ range[1] | Reference |
|---|---|---|---|
| $n_H$ (cm$^{-3}$) | $10^{3.1}$ | - | |
| $T_{spin}$ (K) | 68 | 70-120 | 3, 4 |
| Distance to $\theta^1$ Ori C (pc) | 1.91 | ~1.0 | 5 |
| $S(H\beta)$ (erg s$^{-1}$cm$^{-2}$ arcsec$^{-2}$) | $4.3x10^{-14}$ | $< 5x10^{-12}$ | 4 |
| $S[NII]$ $\lambda$6583 | $7.3x10^{-14}$ | $< 1.49x10^{-13}$ | 6 |
| $N(CI^*)/N(CI)$ | -0.3 | ~ -0.1 | 2 |
| $N(CI^{**})/N(CI)$ | -0.6 | ~ -0.2 | 2 |
| $N(OI^*)/N(OI)$ | -3.0 | -3.5 ± 0.7 | 2 |
| $N(OI^{**})/N(OI)$ | -4.2 | -3.4 ± 0.7 | 2 |
| $N(SiII^*)/N(SiII)$ | -2.2 | -1.4 ± 0.4 | 2 |
| $N(C^0)/N(C^+)$ | -4.0 | > -4.35 | 2 |
| $N(Mg^0)/N(Mg^+)$ | -3.0 | -3.4 ± 0.5 | 2 |
| $N(S^0)/N(S^+)$ | -4.5 | ~ -4.2 ± 1.0 | 2 |
| $N(H_2)$ (cm$^{-2}$) | $1.6x10^{15}$ | $< 3.5x10^{17}$ | 7 |
| $N(H^+)$ (cm$^{-2}$) | $1.2x10^{20}$ | unknown | |

2 Ratio's are in logarithmic units

3 Shuping and Snow, 1997

4 van der Werf & Goss, 1989

5 Baldwin et al. 1991

6 O'Dell & Yusef-Zadeh 2000

7 Savage et al. 1977

Table 2

Log of Column Densities for Best-fitting Model

| | Stage of Ionization | | | |
|---|---|---|---|---|
| | I | II | III | IV |
| H | 21.6 | 20.1 | | |
| He | 20.6 | 18.8 | | |
| Li | 7.8 | 11.3 | | |
| Be | | 1.6 | | |
| B | 7.5 | 11.6 | 8.8 | 7.0 |
| C | 14.1 | 18.1 | 15.9 | 11.0 |
| N | 17.4 | 15.8 | 15.3 | 10.5 |
| O | 18.2 | 16.6 | 15.7 | |
| F | 1.6 | | | |
| Ne | 17.4 | 15.9 | 14.3 | |
| Na | 12.8 | 15.1 | 10.3 | |
| Mg | 13.1 | 16.1 | 14.4 | |
| Al | 10.4 | 14.9 | 12.8 | 11.4 |
| Si | 11.4 | 16.2 | 14.4 | 12. |
| P | 10.5 | 14.8 | 13.0 | 10.8 |
| S | 12.1 | 16.6 | 14.9 | 12.2 |
| Cl | 10.4 | 14.6 | 12.7 | 9.4 |
| Ar | 16.1 | 14.3 | 14.3 | 10.9 |
| K | 11.3 | 13.6 | 11.5 | 7.3 |
| Ca | 8.7 | 11.8 | 13.9 | 7.5 |
| Sc | | 0.2 | 1.6 | |
| Ti | 8.5 | 12.3 | 11.8 | 9.8 |
| V | 8.3 | 11.6 | 10.0 | 8.8 |
| Cr | 9.5 | 13.6 | 12.0 | 10.6 |
| Mn | 11.4 | 14.0 | 12.4 | 10.8 |
| Fe | 12.4 | 16.1 | 14.4 | 13.6 |
| Co | | 1.6 | | |
| Ni | 11.5 | 14.6 | 12.9 | 11.3 |
| Cu | 9.1 | 12.8 | 11.2 | 9.3 |
| Zn | 11.0 | 13.9 | 12.0 | 10.0 |

Table 3

Variation of predicted $H_2$ column density with treatment of destruction rate

| Destruction Rate Treatment | $N(H_2)$ $cm^{-2}$ |
|---|---|
| TH85 | $1.5x10^{19}$ |
| BD96 | $5.0x10^{14}$ |
| Cloudy 96 (Complete $H_2$ molecule) | $1.6x10^{15}$ |

## 9 Figures

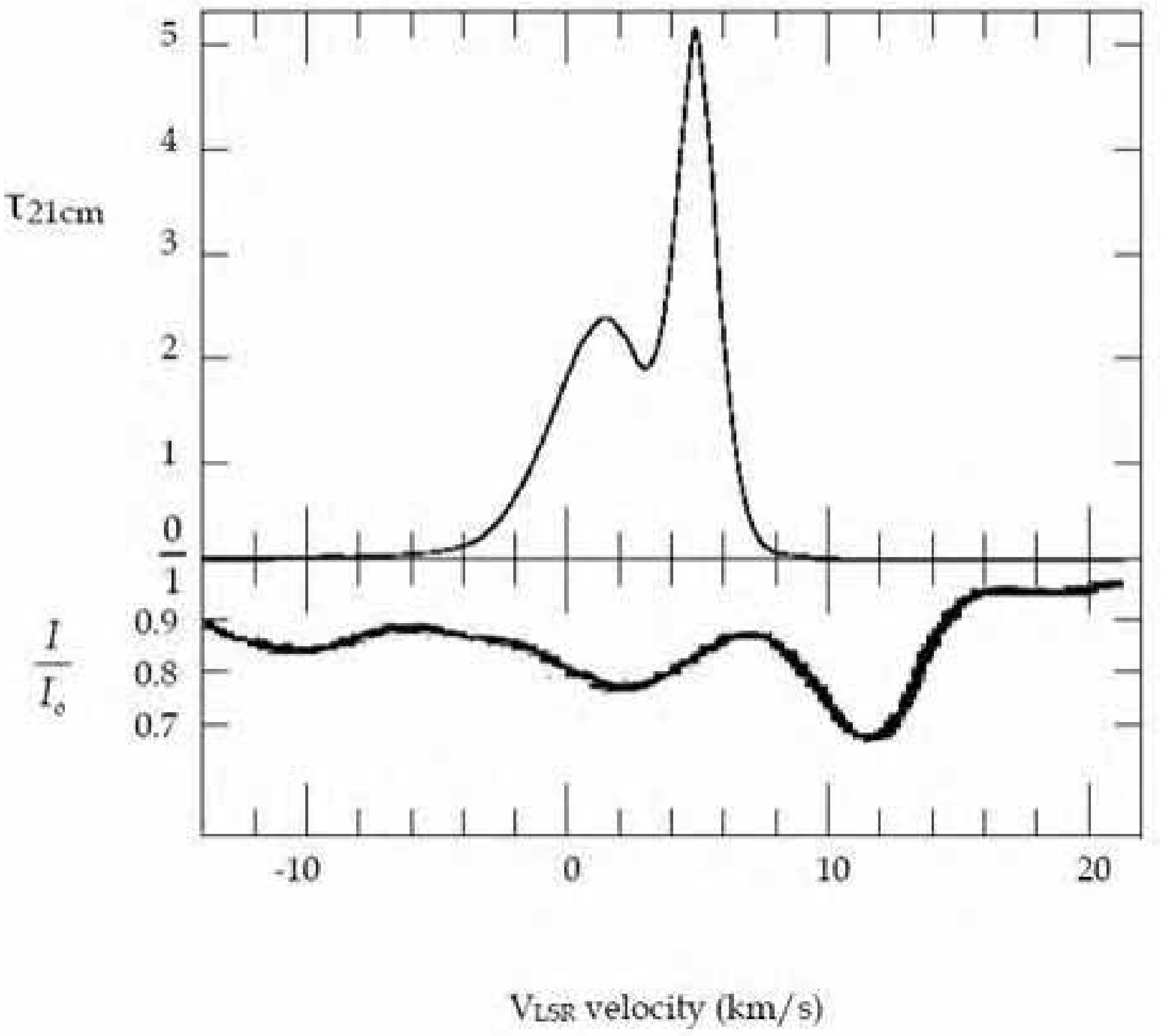

Figure 1 H I 21 cm (upper) and Ca II (lower) profiles observed along the line of sight to central regions of Orion. The 21 cm profile has optical depth on the y-axis, while the Ca II profile is in units relative to the continuum. Note that there does not seem to be a strong correlation between the two profiles, implying that $Ca^{+}$ is not associated with the bulk of the material in the veil.

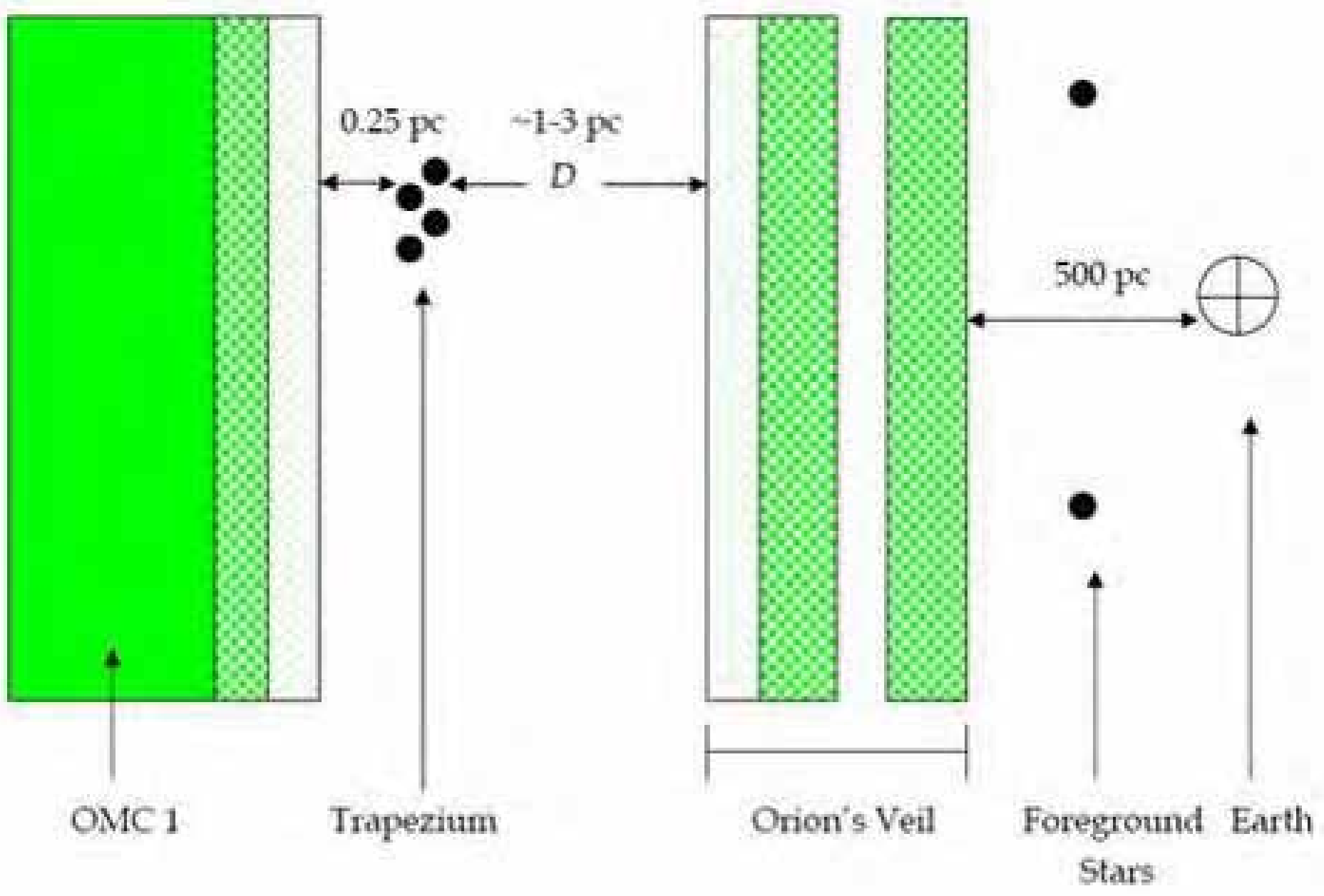

Figure 2 The geometry of the central regions of the Orion Nebula and the Veil. For this figure, solid regions correspond to $H_2$, checkered regions to $H^0$, and dotted regions to $H^+$. Starting with the side of the Veil facing the Trapezium, we have a small $H^+$ zone followed by a PDR. Because of the strong UV flux coming from the Trapezium cluster, $H_2$ never readily forms but instead we have a thick region of atomic hydrogen. The segmented regions in the veil represent the multiple components seen in the 21 cm optical depth profile in Figure 1. The distance *D* represents the distance from the side of the veil facing the trapezium to the trapezium itself, and is the same distance used in all subsequent contour plots.

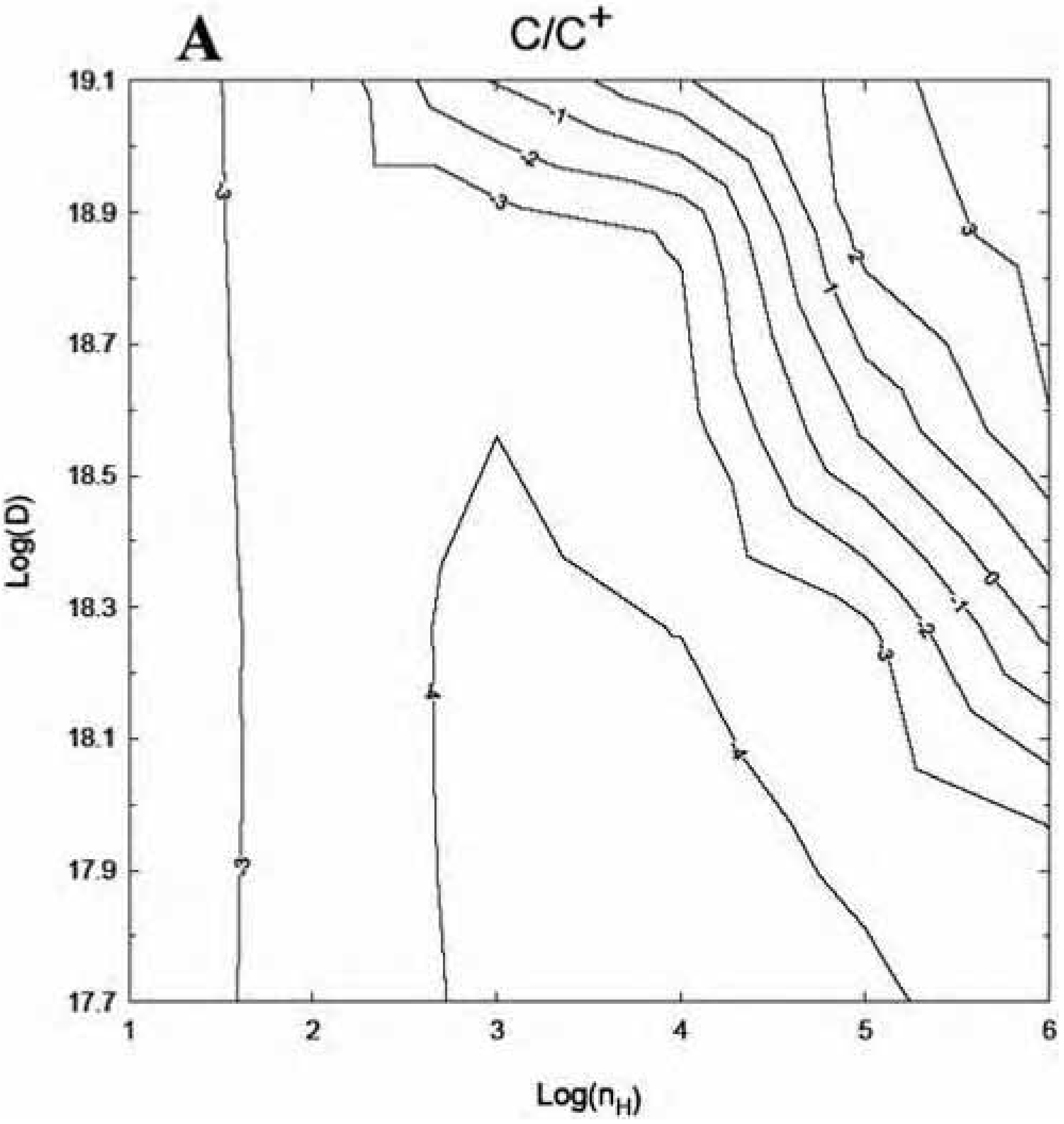

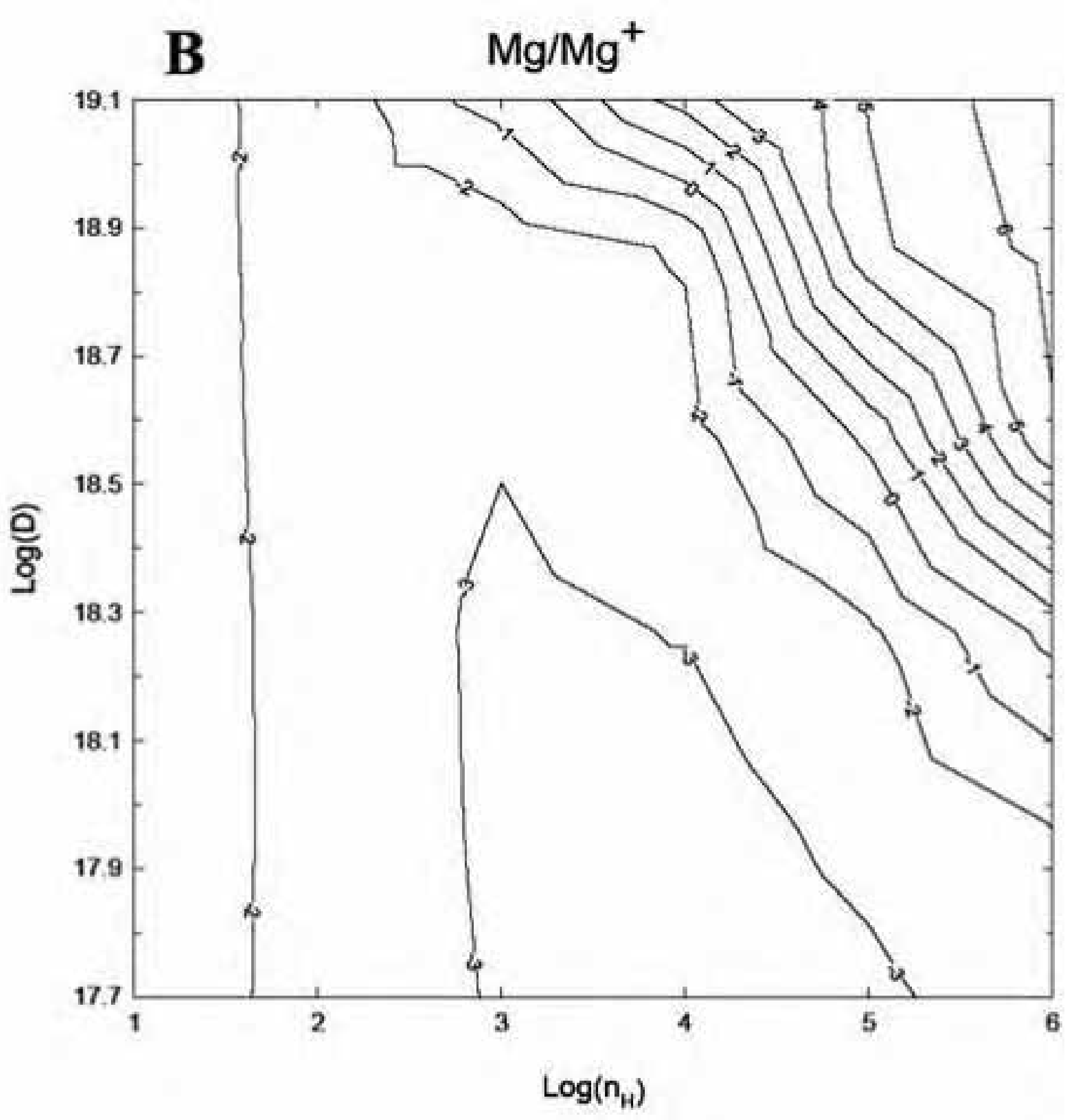

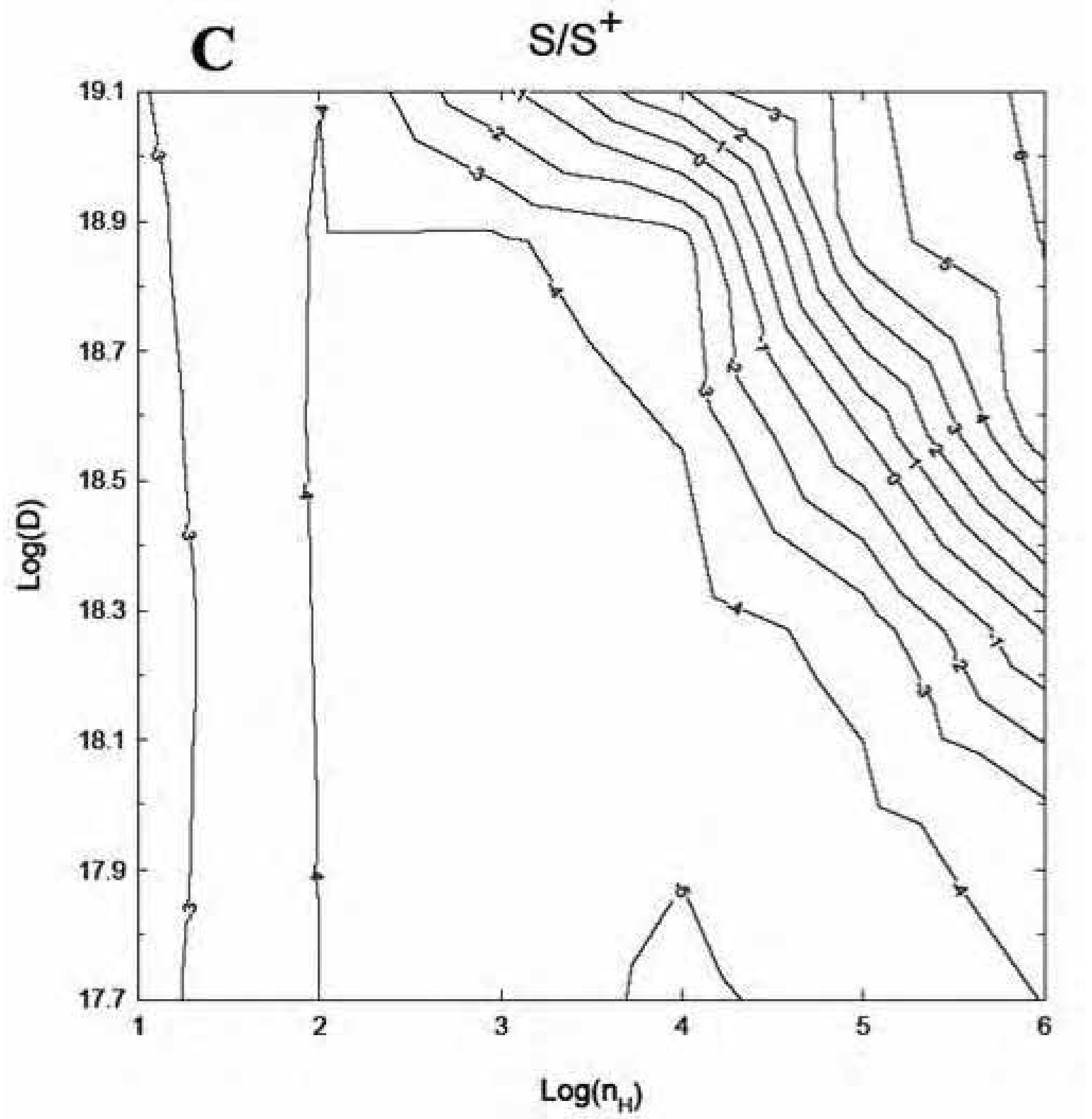

Figure 3 The log of the ratio of atomic to ionized C (A), Mg (B), and S (C), which correspond to the column densities measured in more than one stage of ionization by SS97. The observed 1 σ ratios of C (>-4.35), Mg (-3.4 ± 0.5), and S (-4.2 ± 1.0) agree with most of our calculations. In this and all subsequent contour plots the parameter $n_H = n(H^+) + n(H^0) + 2n(H_2)$ (cm$^{-3}$), the total hydrogen density in all forms.

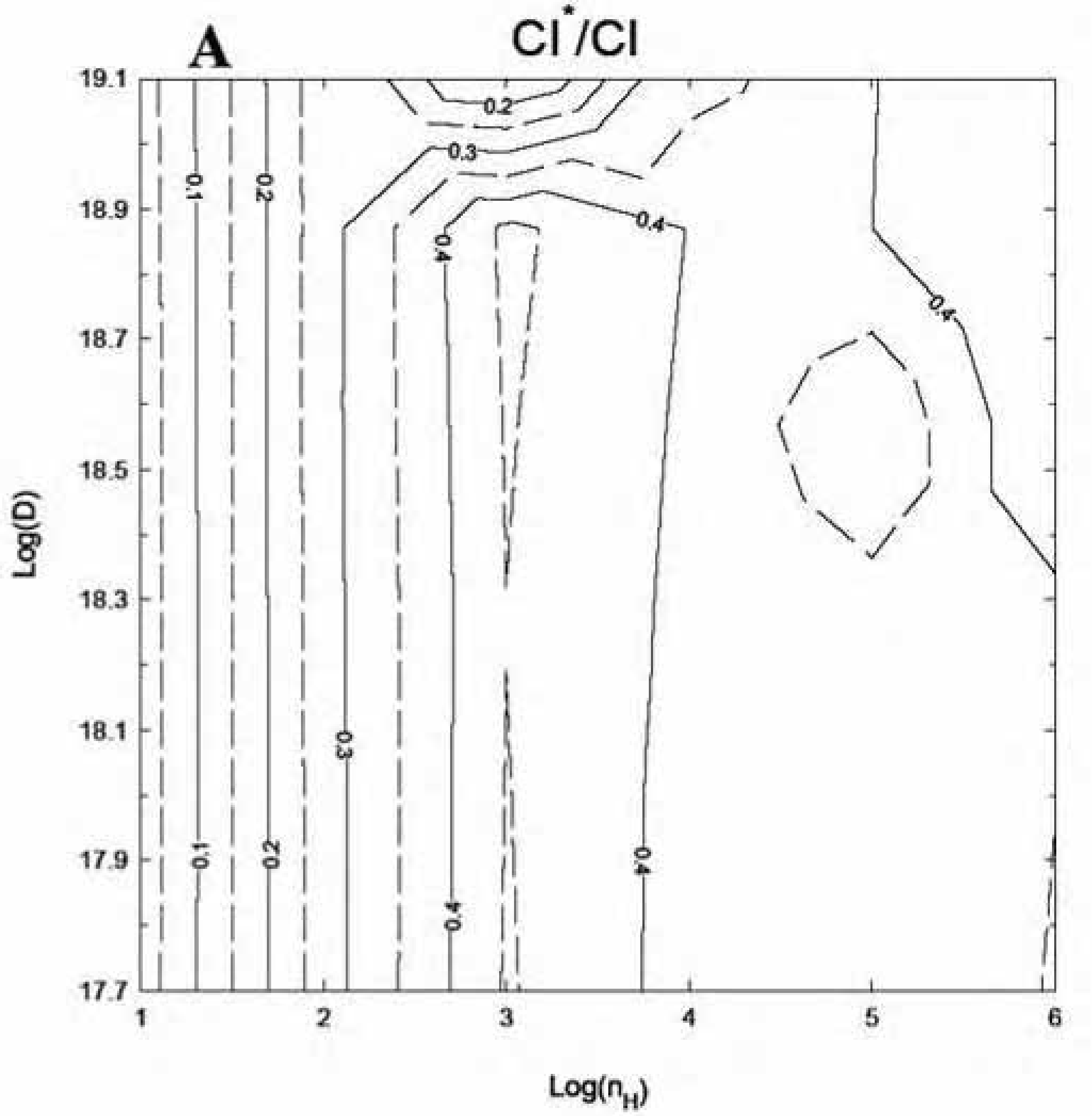

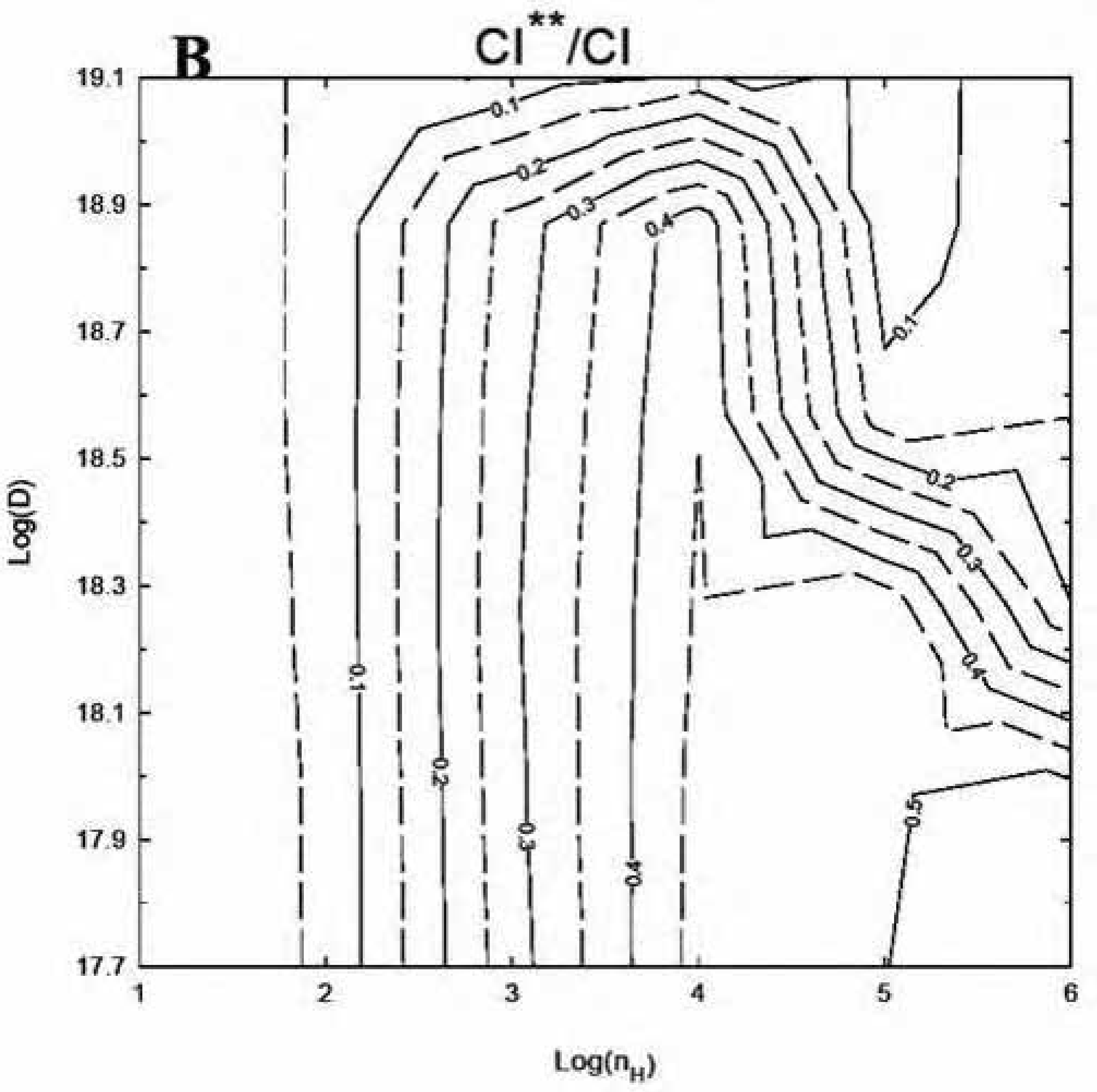

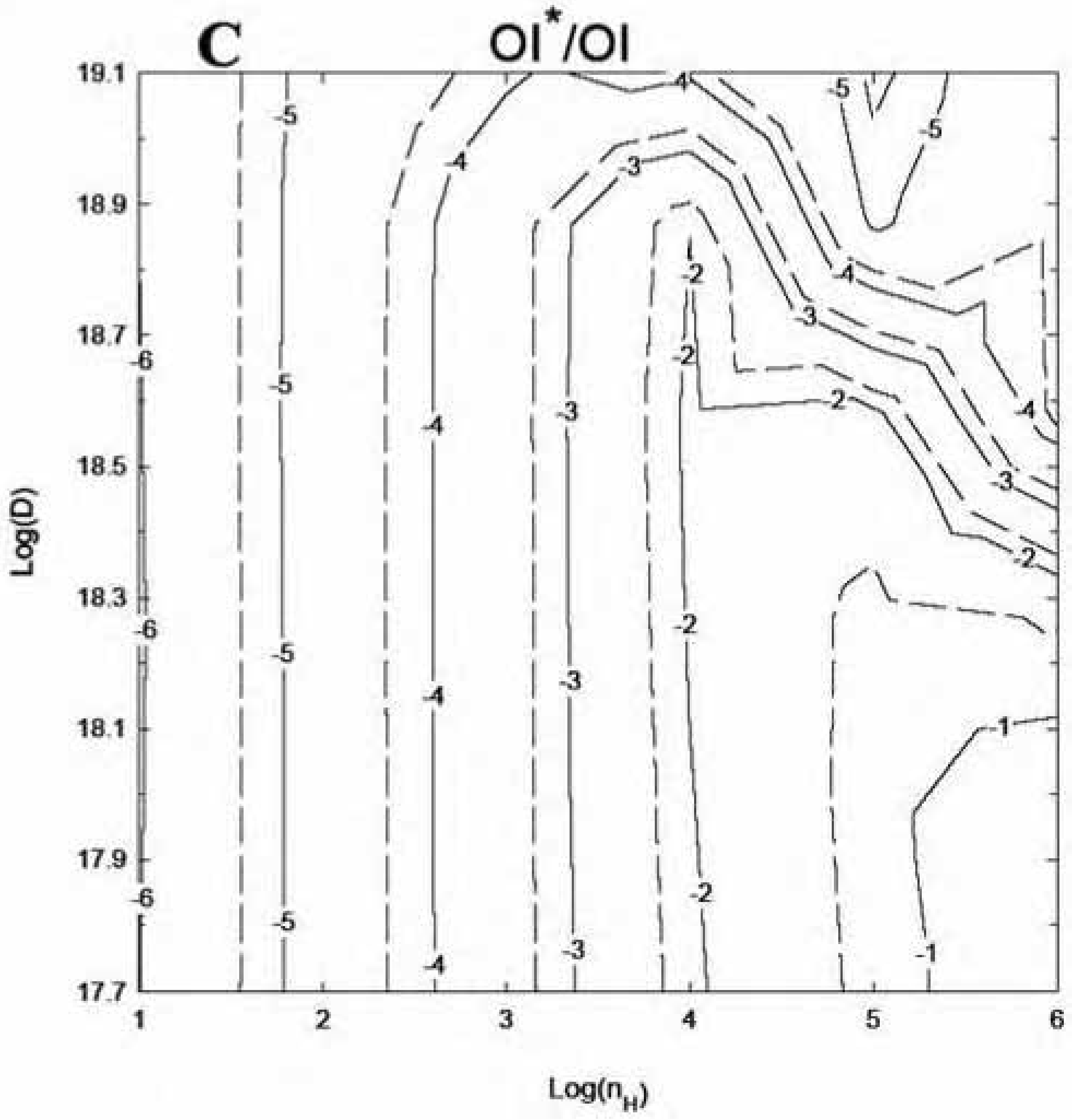

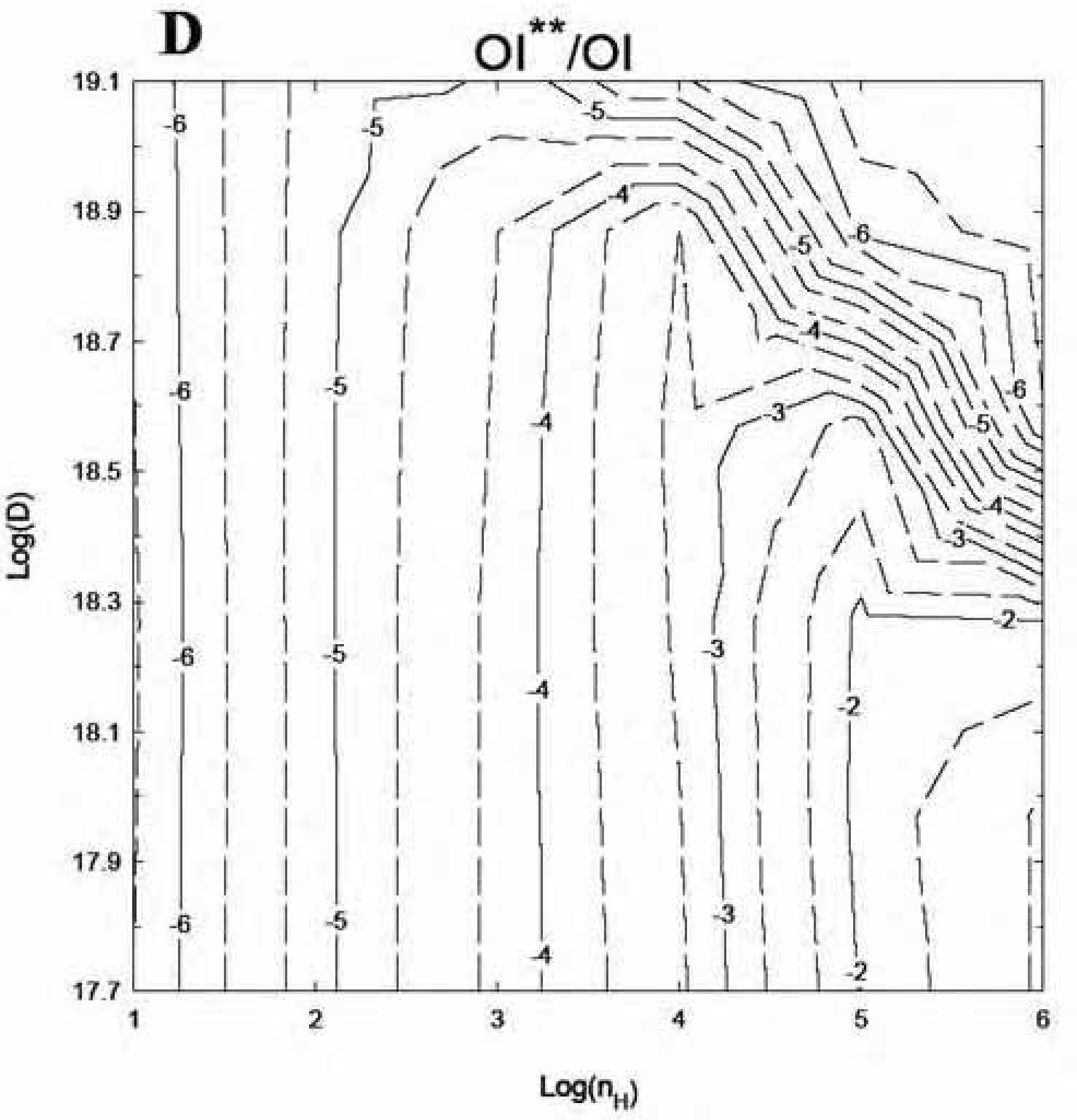

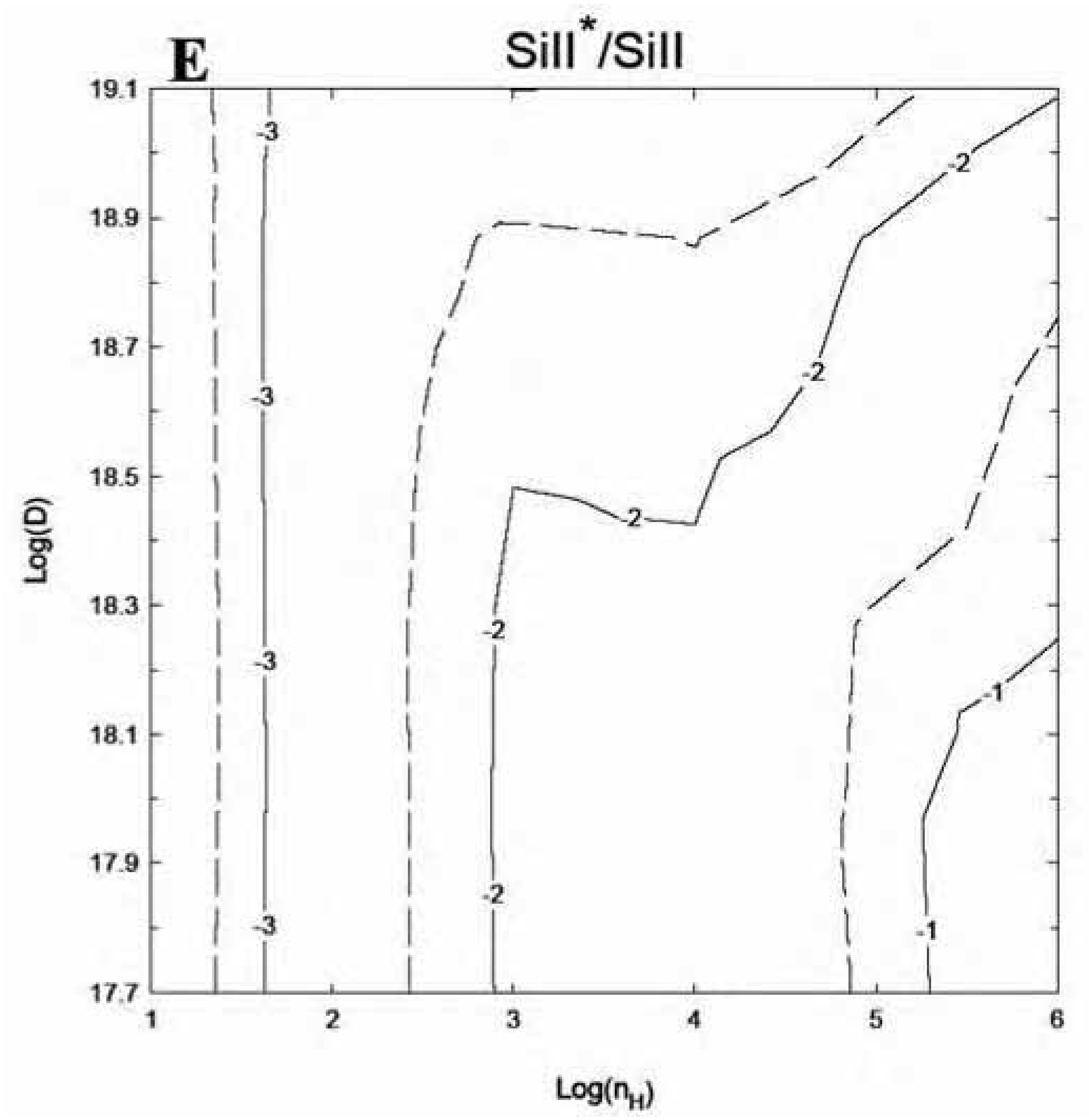

Figure 4 Ratio of excited state column densities of carbon (A & B), oxygen (C & D), and $Si^+$ (E) relative to the ground state. The observed values, with 1σ ranges when available, are: $N(CI^*)/N(CI)$ ~ (-0.1), $N(CI^{**})/N(CI)$ ~ (-0.2), $N(OI^*)/N(OI)$ (-3.5 ± 0.7), $N(OI^{**})/N(OI)$ (-3.4 ± 0.7), and $N(SiII^*)/N(SiII)$ ( -1.4 ± 0.4) The parallel contours show their usefulness as density indicators.

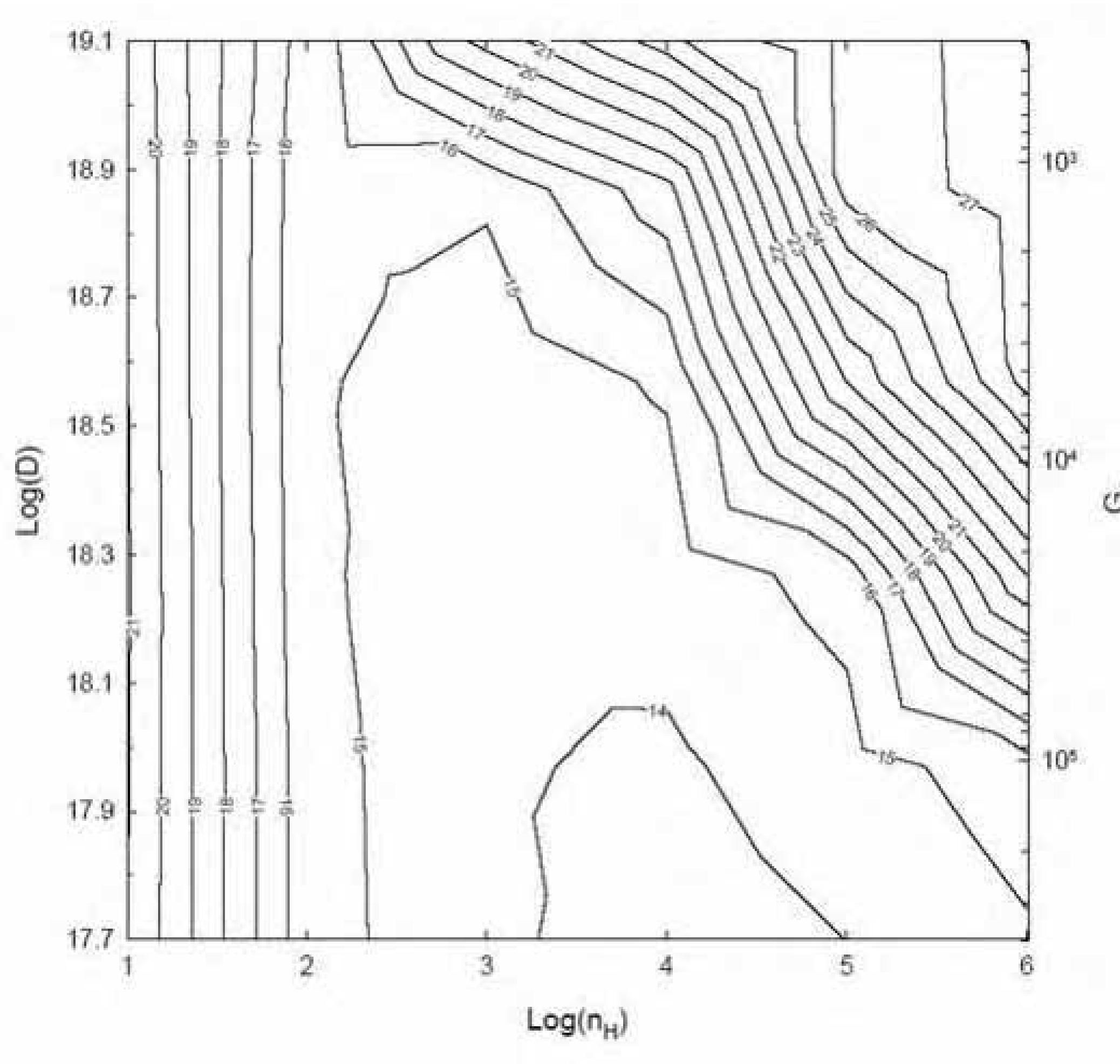

Figure 5 The logarithm of the predicted $H_2$ column density ($cm^{-2}$)is shown. Copernicus observations set an upper limit of $N(H_2) < 3.5x10^{17}$ $cm^{-2}$

(*Log[$H_2$]* < 17.55). In addition to distance, on the right the scale is in units of the average UV radiation field (Tielens and Hollenbach 1985). For the range of densities that our calculations predict for the veil ($10^3$-$10^4$ $cm^{-3}$), our calculations predict an $H_2$ column density less than the Copernicus upper limit for distances less than ~$10^{19.0}$ cm. For very low densities, the large thickness of the $H^+$ region causes the value of $G_0$ to decrease because of geometrical dilution. By the time the hydrogen becomes atomic at these low densities, $G_0$ has decreased to one Habing for all distances. This is why the contours are vertical at low density.

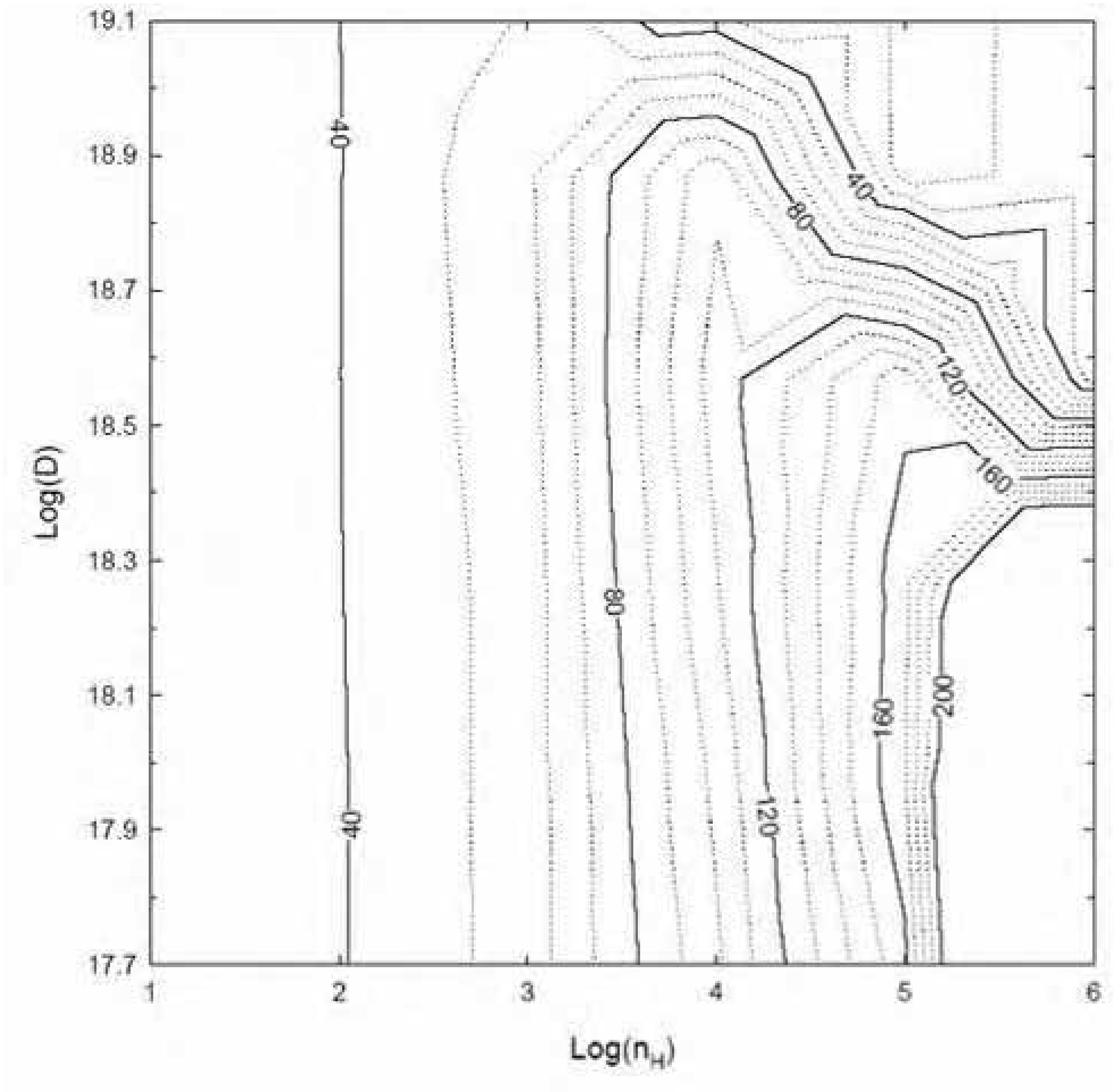

Figure 6 The predicted H I spin temperature (K). For densities less than the critical density of [C II] 158µ ($3x10^3$ $cm^{-3}$), the temperature remains relatively constant. However, for densities greater than this [C II] is no longer an efficient coolant and the heating rate increases with a power of density greater than the cooling. The upper left corner corresponds to calculations that have a large abundance of molecules.

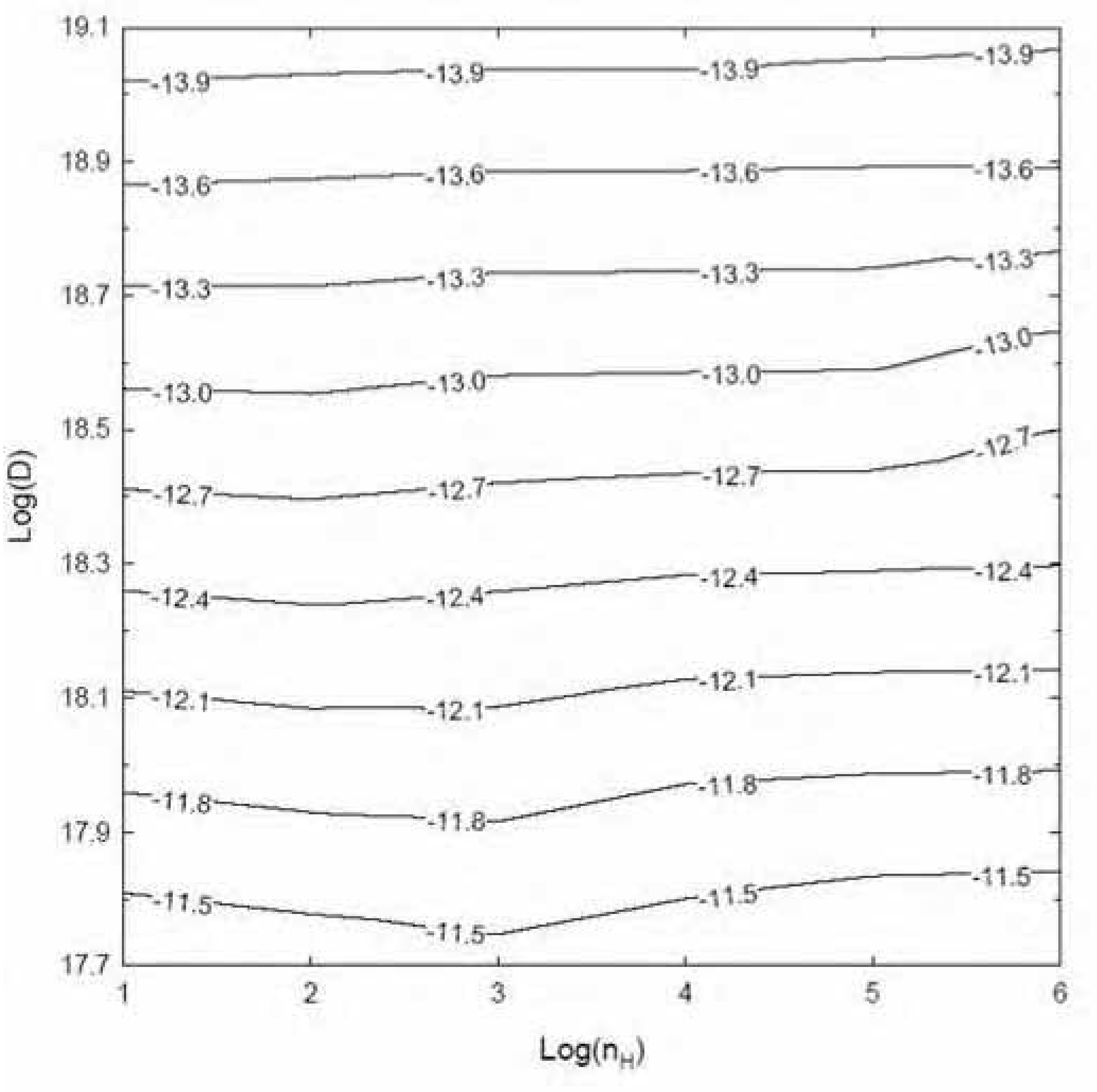

Figure 7 The log of the predicted H surface brightness (ergs $cm^{-2}$ $s^{-1}$ $arcsec^{-2}$ ). The parallel horizontal contours are expected, given the inverse square relationship of surface brightness with distance. The actual surface brightness must be substantially smaller than the total observed surface brightness of $5x10^{-12}$ (*Log[H ]* = -11.3), which mainly comes from the main ionization front.

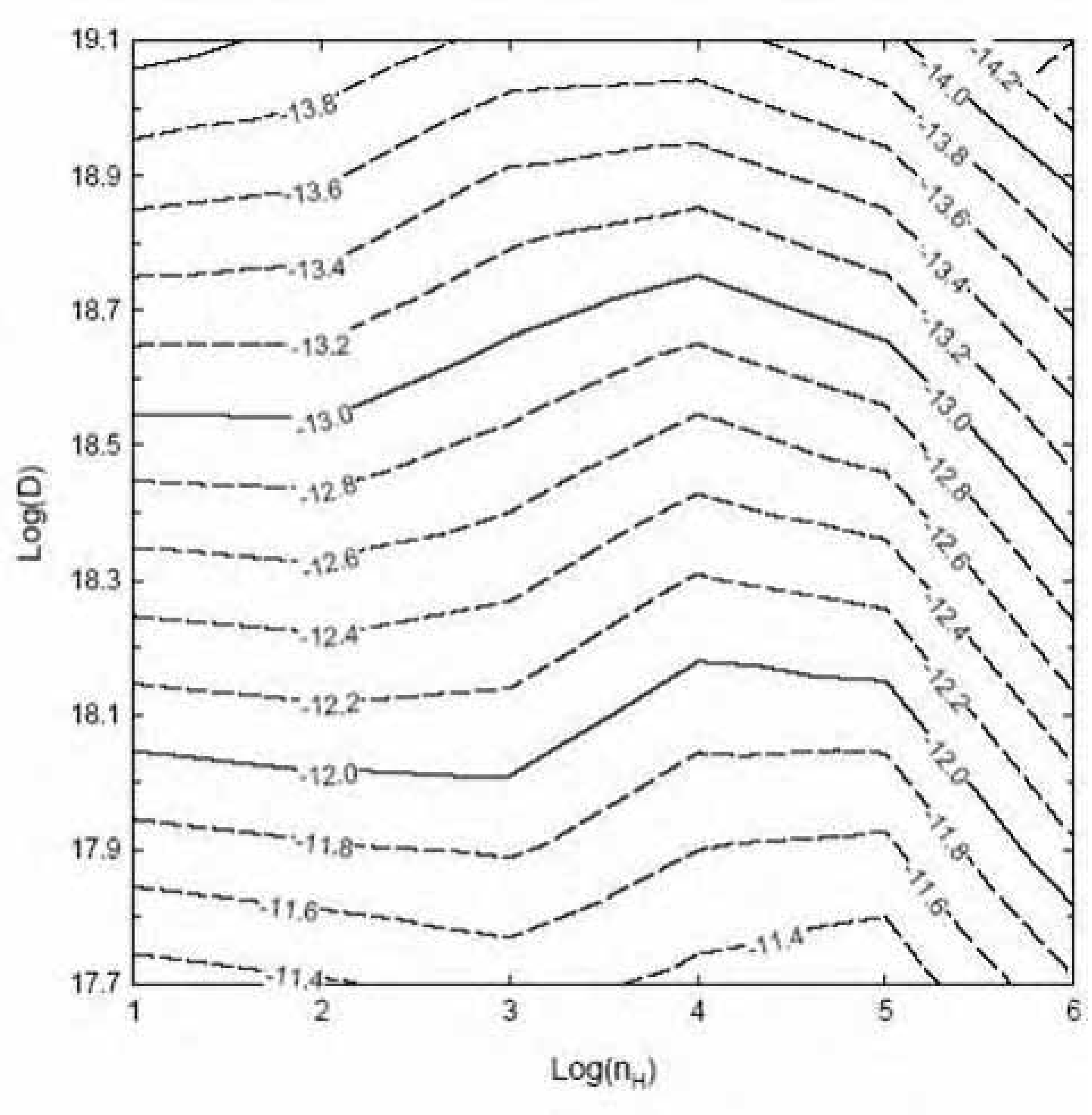

Figure 8 The predicted [N II] surface brightness (ergs $cm^{-2}$ $s^{-1}$ $arcsec^{-2}$ ). Based on calculations by O'Dell, Peimbert, & Peimbert (2003), the upper limit to the veil surface brightness is found to be $1.49x10^{-13}$ ergs $s^{-1}$ $cm^{-2}$ $arcsec^{-2}$ (-12.8). For the range of densities that our calculations predict are most likely for the veil ($10^3$-$10^4$ $cm^{-3}$), this corresponds to a lower limit to the distance of ~$10^{18.5}$ cm.

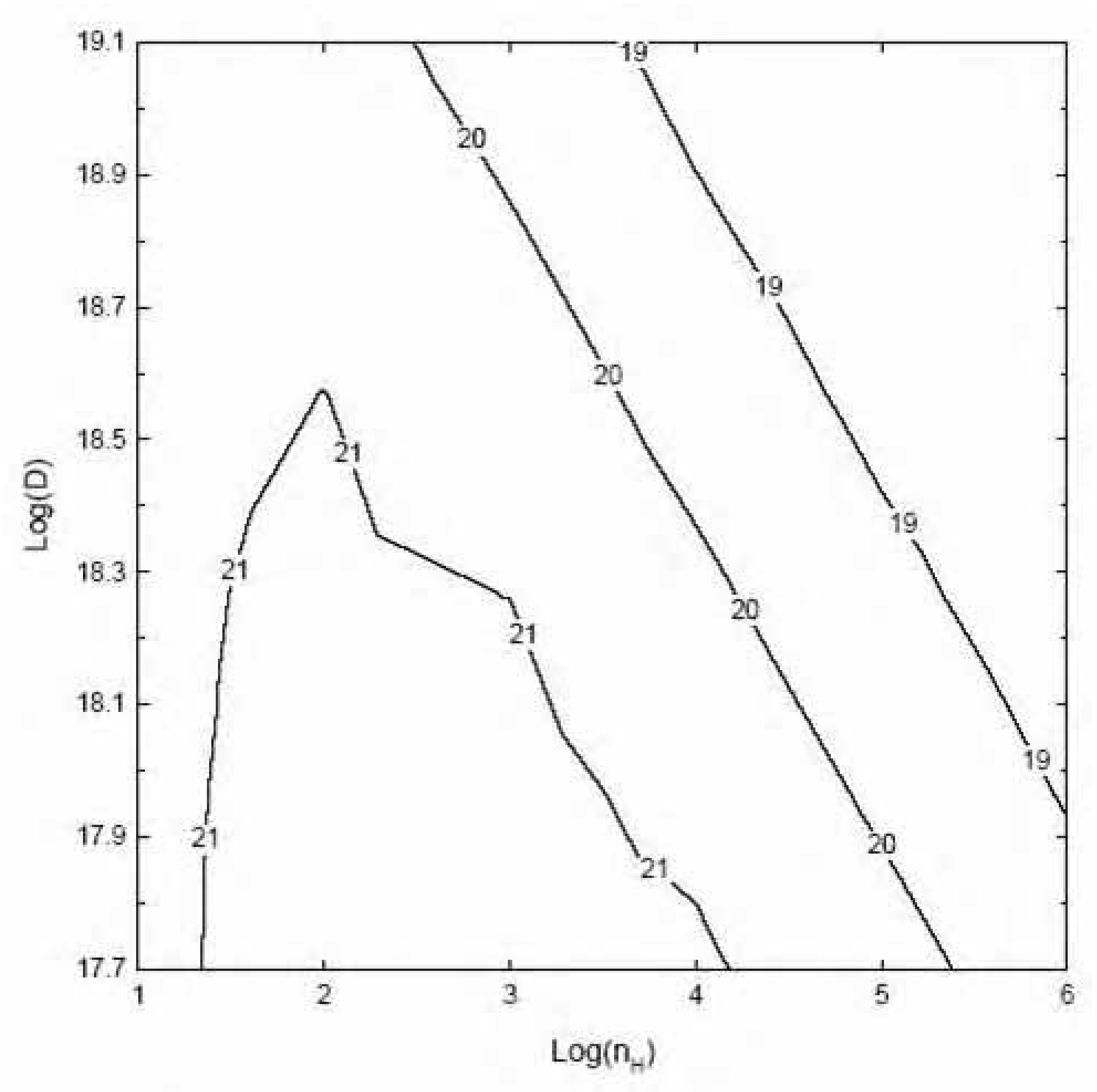

Figure 9 The log of the predicted $H^+$ column density ($cm^{-2}$). This is sensitive to the ionization parameter, which is the dimensionless ratio of flux to density. As expected, larger distances from the ionizing source (and hence smaller flux) and larger densities combine to lower the ionization parameter and hence the H+ column density. For our "best model", about 5% of the total hydrogen is in the form of $H^+$.

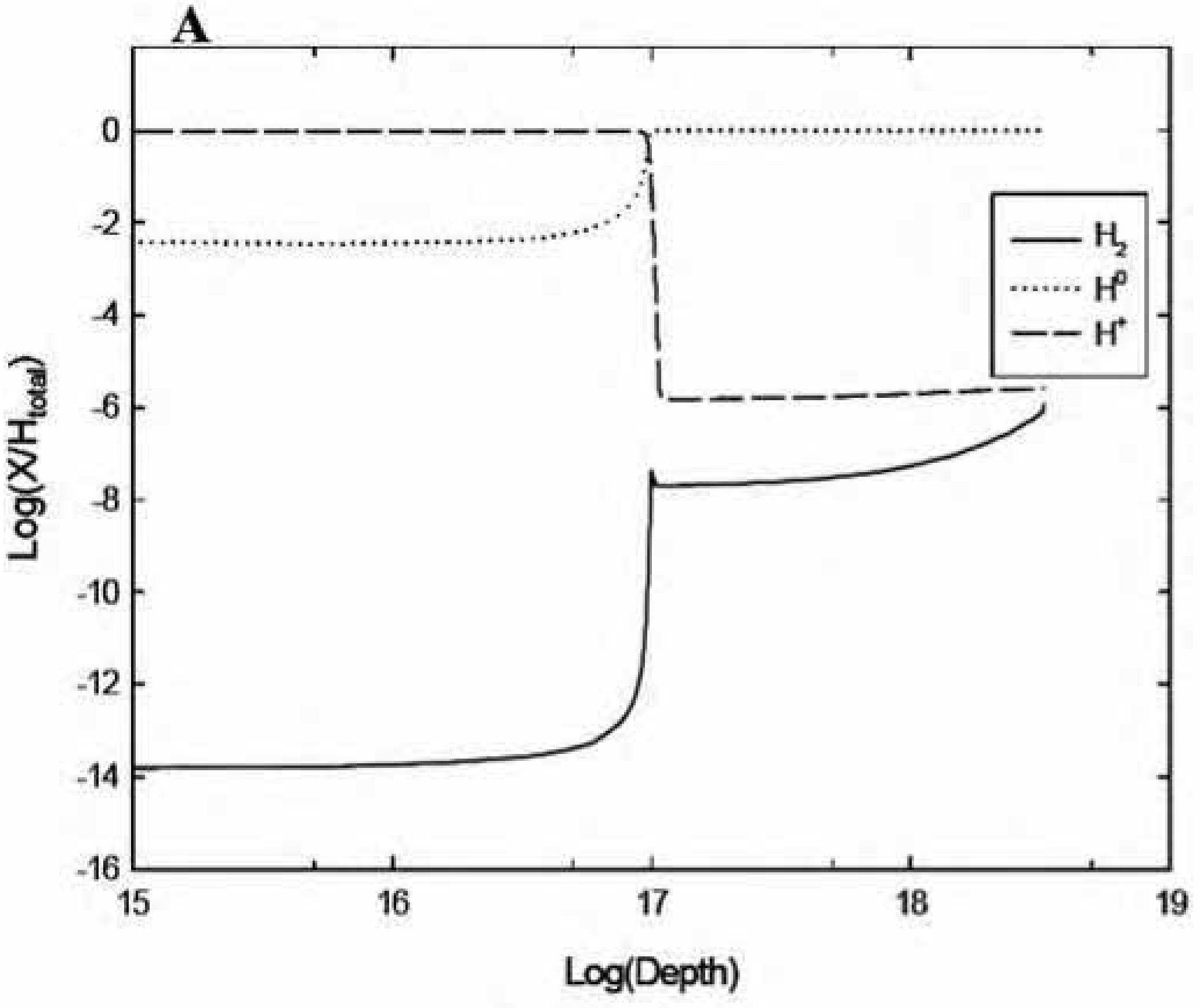

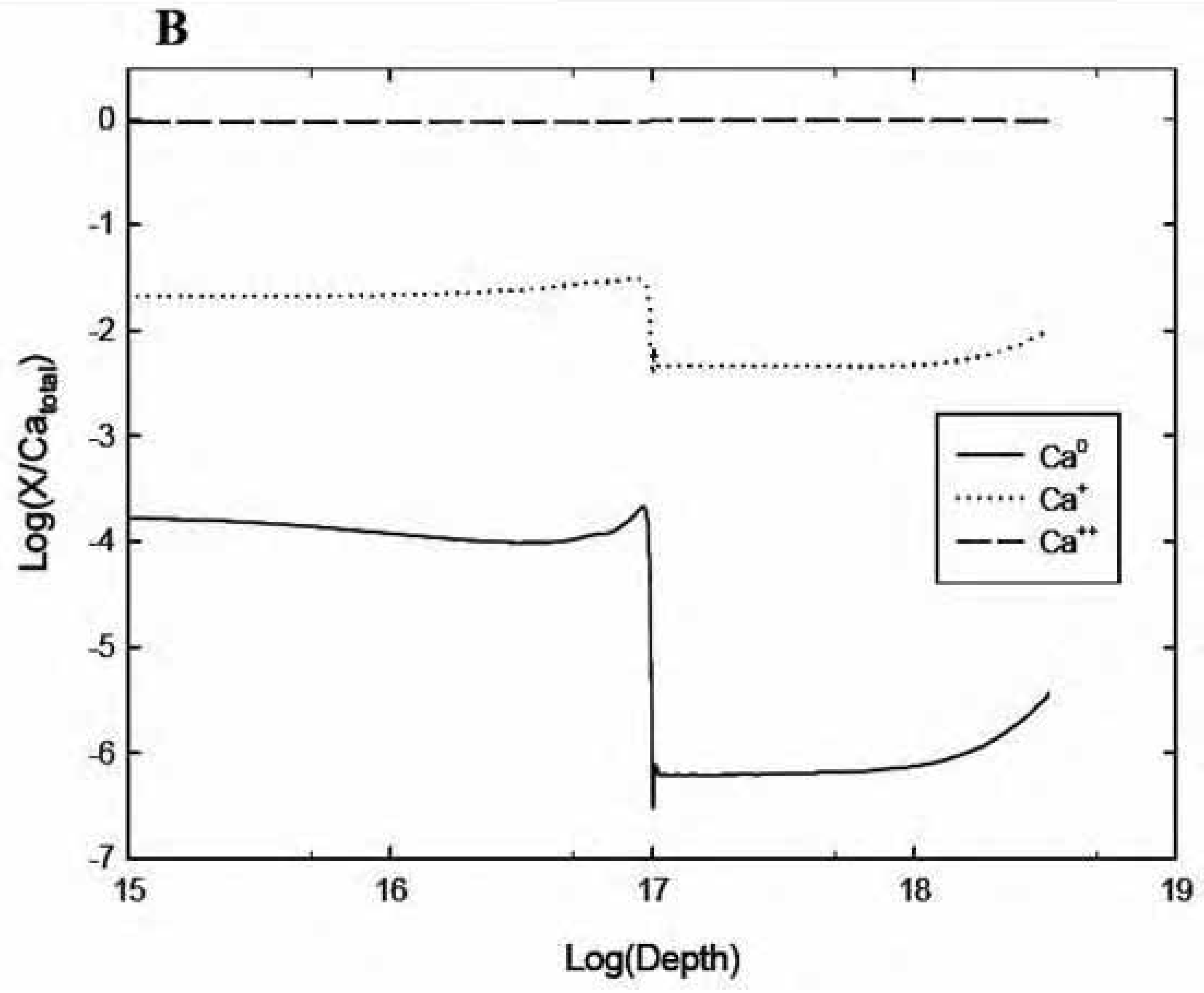

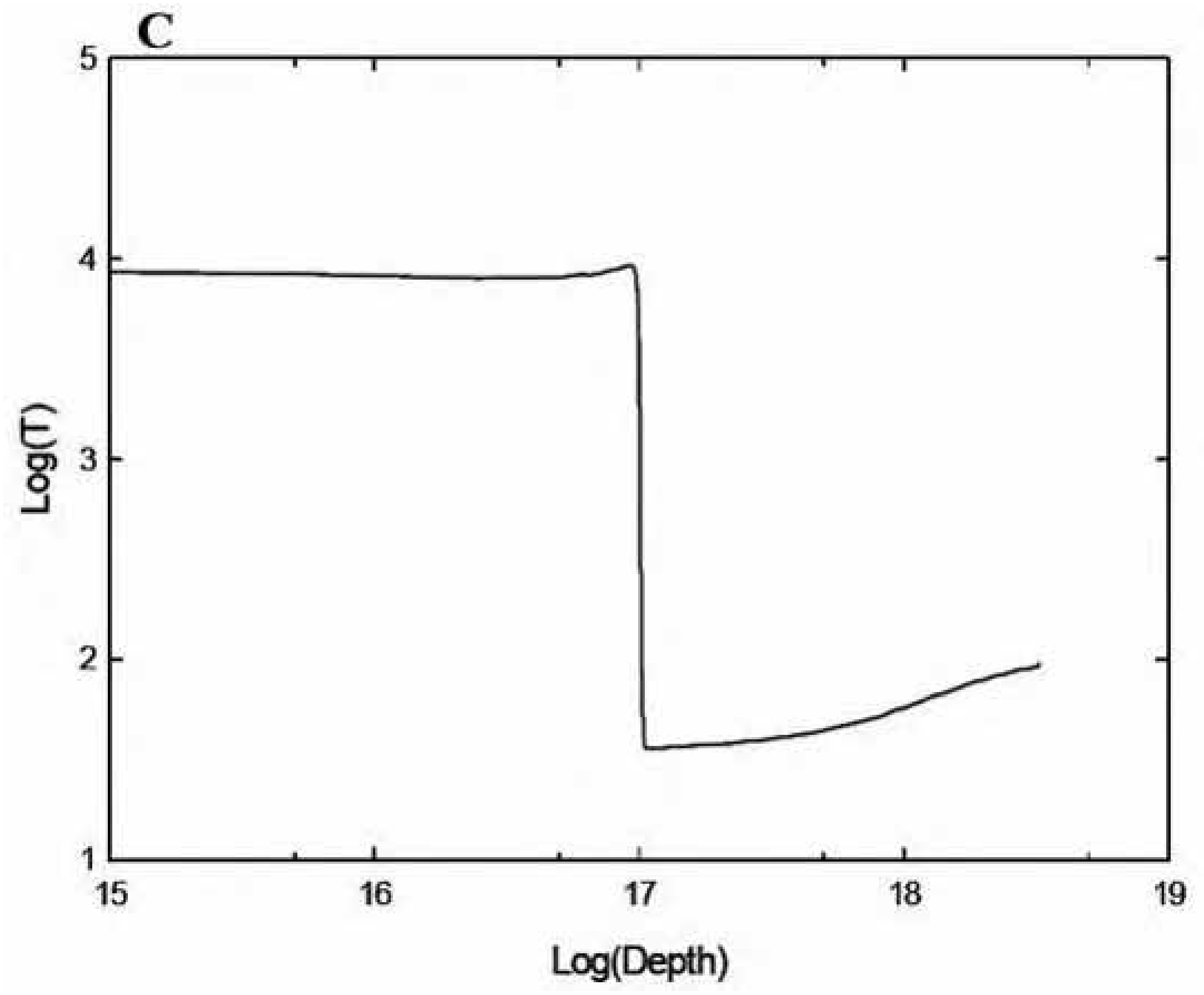

Figure 10 The fractional abundances of hydrogen (A) and calcium (B) in various forms as a function of depth (in centimeters) into the veil for our best model. Also shown is the temperature vs. depth (C) into the veil, with the temperature in units of Kelvins. The drop in temperature corresponds to the hydrogen ionization front. Also of note is that our calculations predict the majority of calcium to be in the form of $Ca^{++}$.

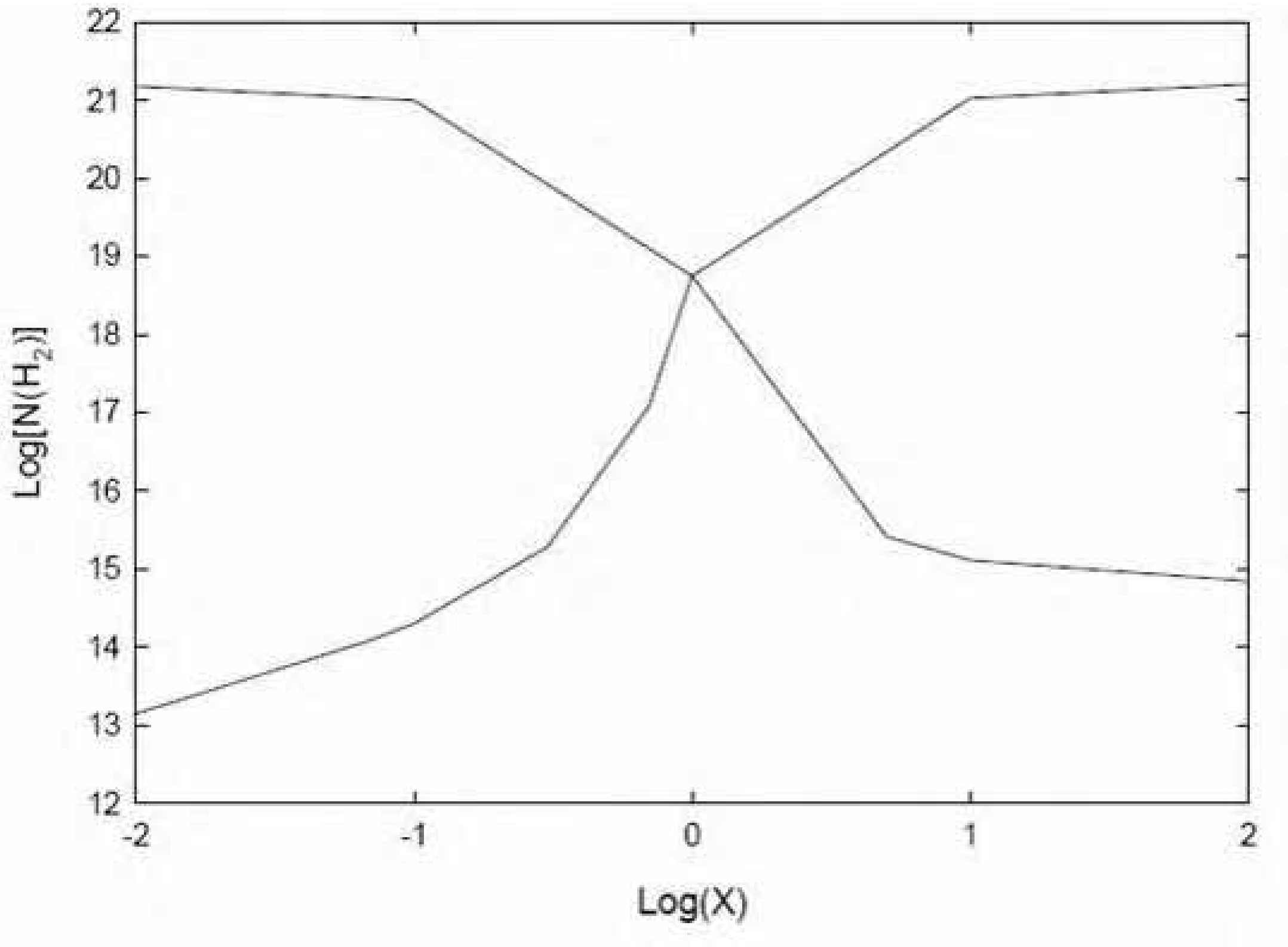

Figure 11 Predicted $H_2$ column density ($cm^{-2}$) as a function of the dissociation and formation rate scaling factor X. This plot clearly illustrates the sensitivity of the $H_2$ column density to the various rates. Changes in either rate of an order of magnitude can cause changes to the $H_2$ column density of up to six orders of magnitude. In order to have confidence in predicting the $H_2$ column density, one must model the $H_2$ formation and destruction processes as accurately as possible, without resorting to approximations.

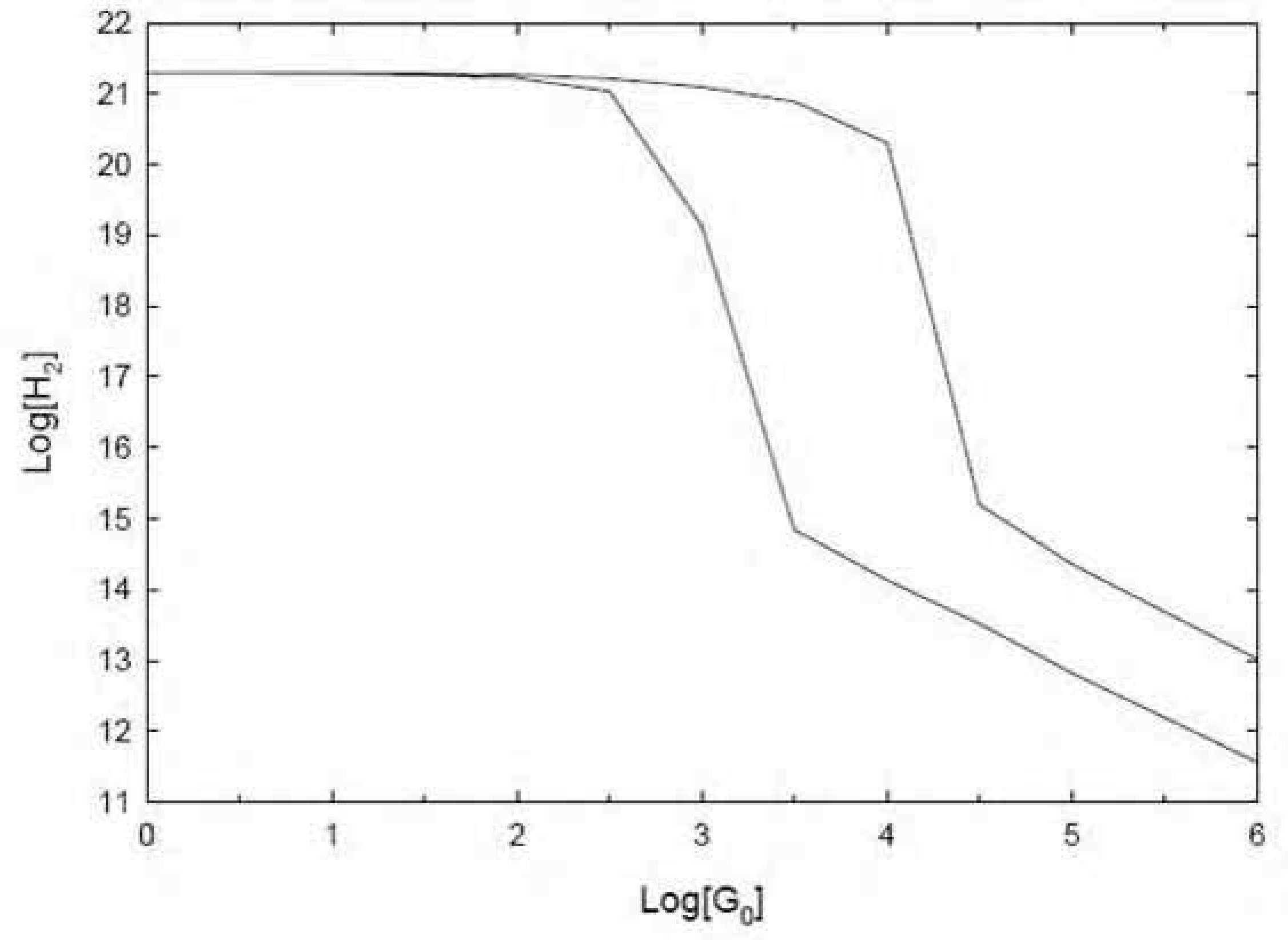

Figure 12 The column density of $H_2$ ($cm^{-2}$) for different grain size distributions over a range of $G_0$. For Log[$G_0$] < 2.5, $H_2$ fully forms for both ISM and Orion grains. For larger values the lack of small grains in the Orion distribution is less effective than ISM grains in shielding the $H_2$ dissociating continuum, and the dissociation rate becomes larger relative to the dissociation rate calculated using ISM grains. This process causes the two size distributions to differ by up to six orders of magnitude for some values of $G_0$.